\documentclass[%
 aip,
 amsmath,amssymb,
 reprint,%
]{revtex4-1}

\usepackage{graphicx}% Include figure files
\usepackage{dcolumn}% Align table columns on decimal point
\usepackage{bm}% bold math
\usepackage[utf8]{inputenc}
\usepackage[T1]{fontenc}
\usepackage{mathptmx}
\usepackage{comment}

\usepackage{multirow}
\usepackage{hyperref}

\newcommand\titlelowercase[1]{\texorpdfstring{\lowercase{#1}}{#1}}

\begin{document}

\preprint{AIP/123-QED}

\title{Guiding the self-assembly of colloidal diamond}

\author{Susana Mar\'in-Aguilar}%
\email{s.marinaguilar@uu.nl}
\affiliation{ 
Soft Condensed Matter, Debye Institute for Nanomaterials Science, Utrecht University, Princetonplein 5, 3584 CC Utrecht, The Netherlands%\\This line break forced with \textbackslash\textbackslash
}%

\author{Fabrizio Camerin}%
\email{f.camerin@uu.nl}
\affiliation{ 
Soft Condensed Matter, Debye Institute for Nanomaterials Science, Utrecht University, Princetonplein 5, 3584 CC Utrecht, The Netherlands%\\This line break forced with \textbackslash\textbackslash
}%

\author{Marjolein Dijkstra}%
\email{m.dijkstra@uu.nl}
\affiliation{ 
Soft Condensed Matter, Debye Institute for Nanomaterials Science, Utrecht University, Princetonplein 5, 3584 CC Utrecht, The Netherlands%\\This line break forced with \textbackslash\textbackslash
}%

\date{\today}% It is always \today, today,

\begin{abstract}
The assembly of colloidal cubic diamond is a challenging process since the shape and interaction parameters and the thermodynamic conditions where this structure is stable are elusive. The simultaneous use of shape-anisotropic particles and strong directional interactions has proven to be a successful path to exclusively nucleate this structure. Here, using molecular dynamics simulations, we explore in detail the conditions where  nucleation of cubic diamond from tetrahedral building blocks is favored. In particular, we focus on the effect of depletion and DNA-mediated interactions to form and stabilize this cubic diamond crystal. We find that a particular balance between the strength and range of the depletion interactions enhances the self-assembly of stable cubic diamond, leading to a narrow region where this structure is nucleated.  Moreover, we determine that stronger short-range depletion attractions may arrest the system leading to the formation of percolating diamond networks or  fully disordered gel structures. Accordingly, the internal arrangements of these structures exhibit a distinct variation in terms of  fractal dimension and the presence of six-membered rings that increasingly acquire internal strain as the arrest gets more pronounced.
With these results we provide a clear route for the self-assembly of cubic colloidal diamond, towards the realization of crystals with superior photonic properties.
\end{abstract}

\maketitle

\section{\label{sec:intro}Introduction}

\begin{figure}[b]
\includegraphics[width=1\linewidth]{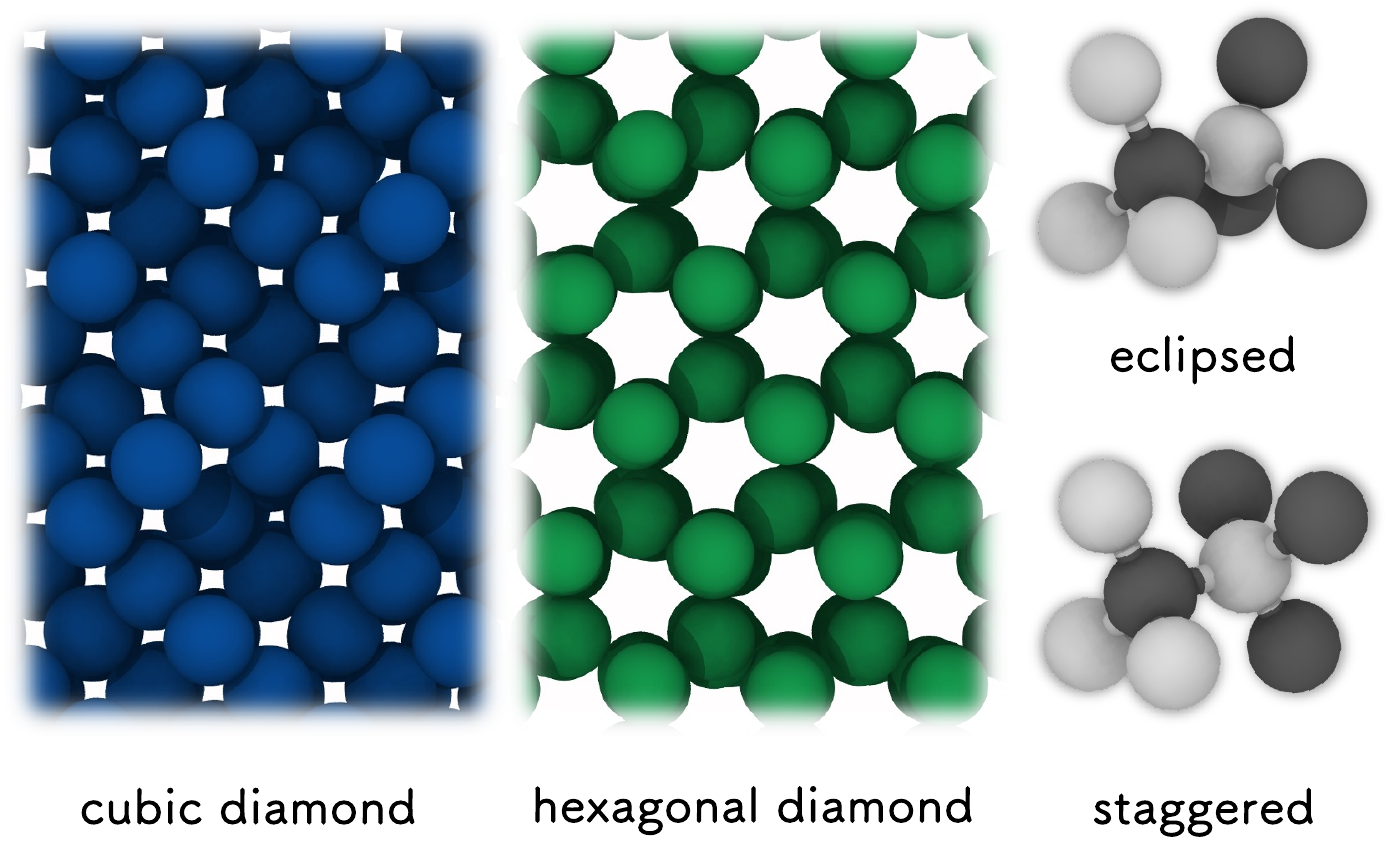} 
\caption{Cubic (left) and hexagonal (middle) diamond structures. In the cubic diamond, the first four nearest neighbors are characterized by  staggered conformations (bottom right) while in the hexagonal diamond three nearest neighbours are characterized by a staggered conformation and one by an eclipsed arrangement (top right).}
\label{fig:cubichex}
\end{figure}

The formation of colloidal crystals, intended as periodic arrays of colloidal particles, is made possible by the ability to synthesize and self-assemble these typical submicrometer-sized building blocks. In contrast to atomic units in macromolecules, the real potential of colloidal particles lies in the ability to precisely control both their shape~\cite{sacanna2011shape, liu2020tunable} and the way they interact,~\cite{li2020colloidal} giving rise to structures and materials with a wide variety of properties and features.~\cite{manoharan2015colloidal}
In this sense, a key role is played by the building blocks that can form so-called open crystals which, compared to close-packed structures, find promising use in photonic crystals in which the transmission of visible light can be controlled.~\cite{li2011colloidal,zhang2009self,ducrot2017colloidal} In this context, the cubic diamond structure is typically taken as the fundamental reference, since it features a complete photonic band gap at low dielectric constants.~\cite{maldovan2004diamond} 
Its structure is characterized by the presence of particles with exactly four bonds in a tetrahedral arrangement. Moreover, these bonds are characterized for being in a staggered chair-like conformation, so that the angle between two bonded particles is $109.5^{\circ}$ as shown in  Fig.~\ref{fig:cubichex}.~\cite{brown2014introduction}

Being able to reproduce this type of structure is therefore of great significance, and over the years several studies have been proposed to understand how to adapt the colloidal particles in a way that  self-assembly of the diamond crystal is promoted. Since low-density crystalline phases are stabilized by anisotropic attractive interactions, patchy particles are ideal candidates for assembling such structures.~\cite{bianchi2011patchy} In fact, when four patches are tetrahedrally arranged, the basic units of the diamond conformation could be stabilized.~\cite{romano2010phase,noya2010stability,romano2011crystallization} Although it is  possible to assemble cubic diamond under certain conditions, often other structures with a similar (free) energy but lacking a photonic band gap, are also obtained. Among these is the hexagonal diamond,~\cite{romano2011crystallization,leoni2021nonclassical,doppelbauer2012competing} which is characterized by three tetrahedral units in a staggered conformation and one in an eclipsed  conformation as illustrated in  Fig.~\ref{fig:cubichex}. Moreover, the size of the patch appears to play a crucial role. For instance, smaller patches are prone to stabilize complex clathrate arrangements.~\cite{noya2019assembly} On the other hand, when patches are too wide spontaneous crystallization could be hampered.~\cite{romano2011crystallization} Other recent approaches based on patchy particles rely on inverse designing the single units depending on the target structure that has to be assembled.~\cite{romano2020designing, ma2021inverse,ma2019inverse, patra2018programmable,rovigatti2022simple} In order to introduce selectivity in the interactions and to exclude the formation of undesired patterns, a variable number of patches per particle is assigned a certain `color' and species of different relative concentrations can be admitted to the self-assembly procedure. This type of approach based on the use of patchy particles, could find experimental realization in the use of single-stranded DNA,~\cite{tian2020ordered,liu2016diamond} which allows for high selectivity through a clever diversification of the nucleotide sequences.

Since closed ring patterns with an even number of components can be identified in crystalline diamond, the creation of building blocks that favor the formation of such motifs is another avenue to pursue.~\cite{morphew2018programming} Some recent studies lead in this direction. In particular, it has been shown that the use of a binary mixture of tetrahedral patchy particles allows the sole formation of rings with an even number of units, thus avoiding rings with five or seven members that are typical of disordered structures.~\cite{neophytou2021facile} The same idea is applied to a system of triblock patchy rods, where the extremities of the building blocks exhibit different interactions among each other.~\cite{neophytou2021self,rao2020leveraging}
Finally, a promising route for obtaining colloidal diamond relies on polyhedrally shaped nanoparticles. In this sense, few attempts were based on truncated tetrahedral nanoparticles~\cite{damasceno2012crystalline} and lately on modified gyrobifastigiums, whose shape consists of a complex 8-sided polyhedron.~\cite{zhou2022route} For the latter method, an experimental counterpart still has to be designed. 

These attempts and studies all point in the direction of devising a simple route to assemble crystalline diamond, given the potential this structure holds. However, among other difficulties, it is challenging to exclusively favor the formation of cubic diamond and not other polymorphs, as well as the functionalization of colloids to obtain tetrahedral patchy particles is not immediate and easy to implement experimentally. An interesting idea involving the use of tetrahedral colloidal molecules has been presented very recently by He \textit{et al.}~\cite{he2020colloidal} The advantage of using these objects is that they can be built from very simple fundamental units and  can therefore be provided with a very specific shape. Indeed, it was proposed to use colloidal fusion~\cite{gong2017patchy} to form clusters of partially compressed spheres, where four monomers are at the vertices of a tetrahedron and one links them at the center, see Fig.~\ref{fig:model}. The latter remains exposed on the faces of the tetrahedron and when covered with DNA plays the role of a triangular patch. The successful assembly of cubic diamond based on such building blocks has to be attributed primarily to the shape of the colloidal molecules, which also excludes the emergence of other types of lattices. In fact, thanks to the tetrahedral arrangement of the  monomers, a steric interlocking mechanism takes place, allowing for the stabilization of staggered conformations between neighboring particles  as shown in Fig.~\ref{fig:model}. Energy minimization for this kind of conformation is further enhanced by  DNA interactions acting among different units. In addition to this, the presence of DNA fragments and micelles in the experimental system also causes depletion interactions to take place between the colloidal molecules, inducing a further effective attraction. These interactions are far from being only collateral forces between colloidal particles but on the contrary they can be exploited for different purposes.~\cite{miyazaki2022asakura} Their short-range and controlled strength of attraction offers an ideal playground for studying
the glass transition and gelation of dense colloidal suspensions.~\cite{zaccarelli2007colloidal,ruiz2021role, patrick2008direct, kumar2013block}  
For colloidal particles, the addition of depletant agents has been demonstrated to be effective in inducing the assembly of crystalline phases.~\cite{rossi2011cubic, kraft2012surface,liu2020tunable}
With respect to the tetrahedral clusters described above, the role of the depletion interactions is still not entirely clear, particularly in relation to the stronger attractions induced by the presence of DNA. It is expected that depletion promotes the interactions between particles and thus the formation of diamond, but to what extent this occurs remains to be defined. In principle,  there might be conditions in which crystallization is greatly promoted and others in which the formation of ordered structures is hindered. %random or arrested states prevails. 

With the purpose of guiding the assembly of colloidal diamond and determining the most optimal conditions under which this occurs, we here perform extensive molecular dynamics simulations of tetrahedral colloidal molecules, based on the experimental realization introduced above.~\cite{he2020colloidal} In particular, we explore the phase behavior of the system under  different depletion conditions, by changing the depletant size and its concentration in solution, thus exploring a wide range of interaction ranges and strengths. At first, we study this at small colloid densities, where interesting aspects on the formation of crystalline and gel-like structures may emerge, before exploring a wide range of colloid packing fractions. Furthermore, in order to disentangle the contribution of DNA and to investigate  its role in the crystallization of diamonds, we also consider the case in which the central core particle of the cluster is not coated with DNA, thus interacting solely via depletion interactions, similarly to the monomers.

In particular, we observe nucleation of grains of cubic diamond at a relatively narrow range of depletion sizes and for weak depletion interactions.
On the other hand, at large depletant concentrations, the induced attractions are strong enough to arrest the system leading to percolating networks of cubic diamond or to completely random aggregates. 
We then explore the inner and global structural properties of the cubic diamond and the arrested phases, in terms of nucleation paths, fractal dimensions, and the emerging patterns in the respective structures. 

The remainder of this article is organized as follows. In Section~\ref{sec:methods} we discuss the model and the effective potentials employed to mimic the DNA patchy and depletion interactions in the simulations. In Section~\ref{sec:results}, we first explore the effect of changing the size ratio between the core particles and the lobes located at the vertices of the tetrahedron in terms of the number fraction of particles with a diamond arrangement. %This discussion is carried out 
We then perform simulations for colloidal molecules both with and without a DNA patchy interaction between the core particles. Subsequently, we move our attention to the structural properties of all the phases. Finally, we focus on a particular depletant size and explore the full kinetic phase diagram for varying colloid packing fractions. In Section~\ref{sec:conclusion} we summarize the main results and discuss our conclusions.

\section{\label{sec:methods}Methods}
\subsection{Model and Methodology}
\begin{figure}
\includegraphics[width=1\linewidth]{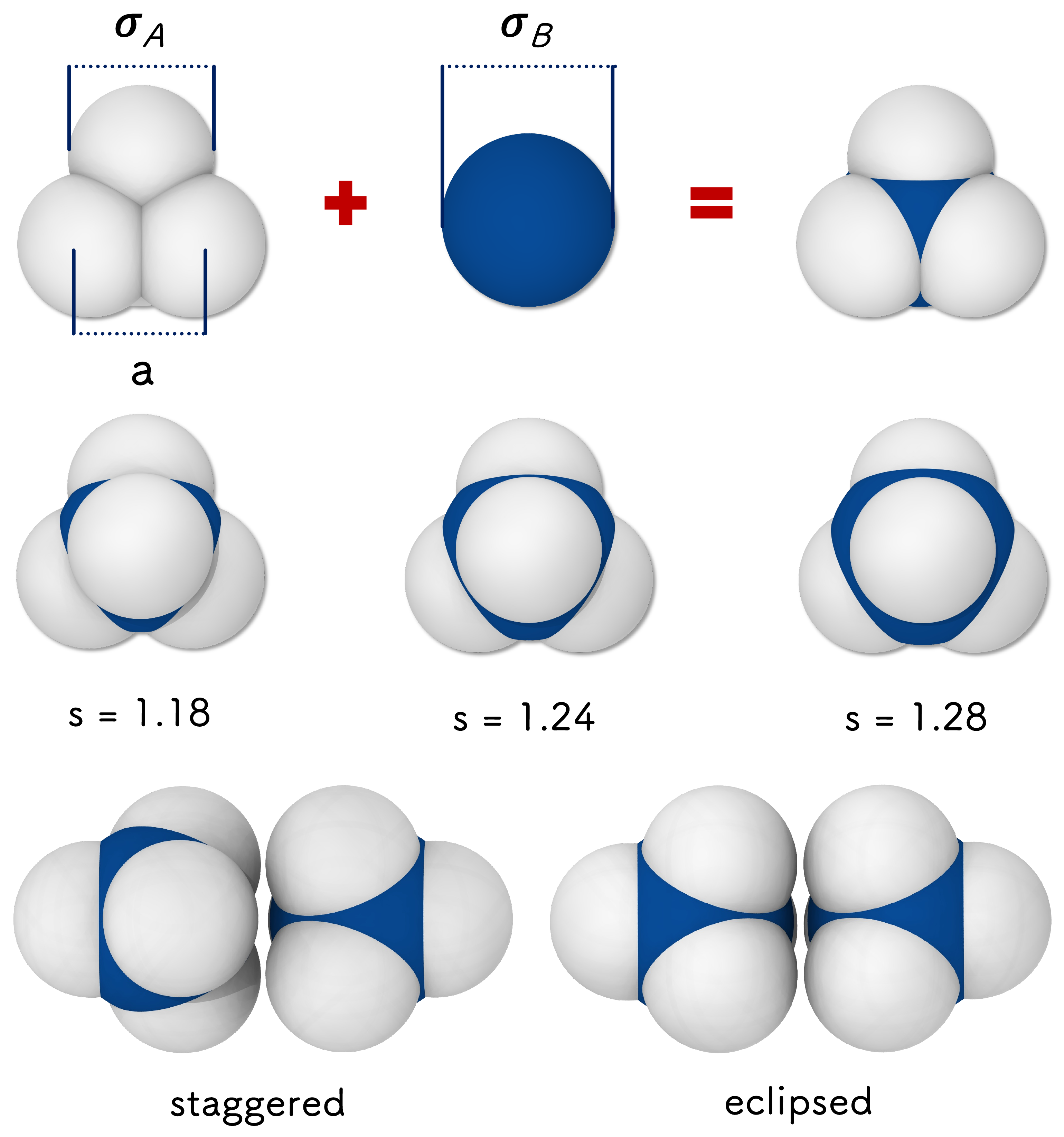} 
\caption{Building blocks for colloidal diamond. Each colloidal molecule is made of four spheres (or lobes) of diameter $\sigma_A$ located at the vertices of a tetrahedron and of a central core particle of diameter $\sigma_B$. In the central row, we show three representative particles for varying core-to-lobe size ratios $s=\sigma_B/\sigma_A$. The DNA-coated central particle forms four disconnected patches for $s=1.18$, while a more uniform coverage is found for $s=1.28$ making the interaction  less  directional. In the bottom row, we show the (preferred) staggered and the (unfavored) eclipsed conformation for such tetrahedral colloidal molecules.}
\label{fig:model}
\end{figure}

We base our model on the experimental realization by He \textit{et al.},~\cite{he2020colloidal} where anisotropic colloidal molecules with a tetrahedral shape and directional interactions were used. Such particles are formed by four colloidal spheres of diameter $\sigma_A$ located at the vertices of a tetrahedron of side $a$ and by a central spherical particle of diameter $\sigma_B$. We take $\sigma_A$ as the unit of length in our simulations. 
The extent to which the central and the satellite spheres are allowed to overlap with each other is controlled both by their core-to-lobe size ratio $s=\sigma_B/\sigma_A$ and by the tetrahedron side length $a$.
Schematic pictures displaying the structure of the colloidal particles at play are presented in Fig.~\ref{fig:model}. From the conformation of a single colloidal molecule, it is also clear how the interlocking mechanism between two tetrahedral colloidal clusters favors staggered conformations, with one lobe of a molecule interacting with two lobes of the second one. In this way, eclipsed arrangements are disfavored.

\begin{figure}
\includegraphics[width=1\linewidth]{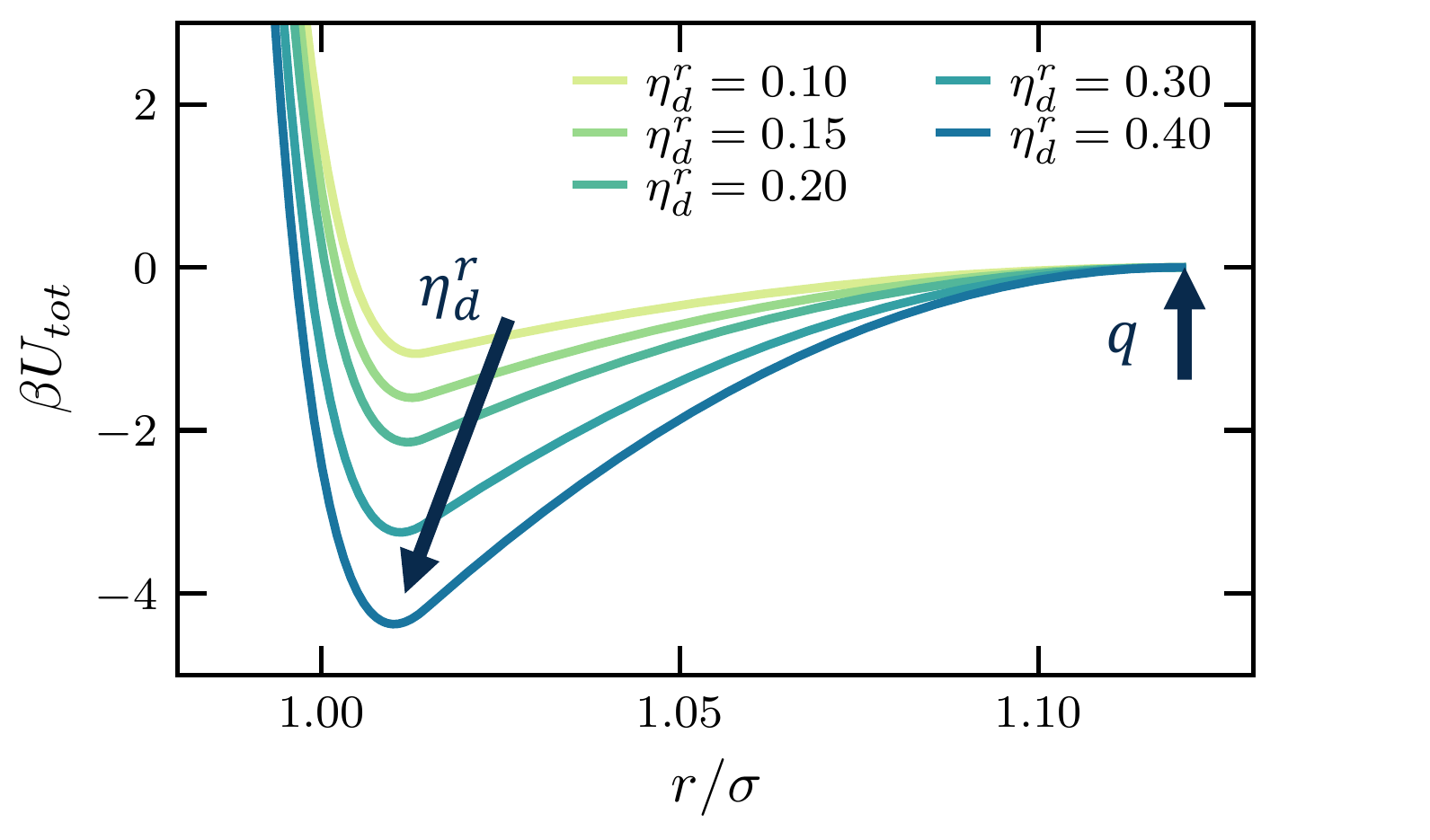} 
\caption{Total pair potential  $U_{tot}(r)=U_{Mie}(r) + U_{AO}(r)$ for different depletant reservoir packing fractions $\eta_d^r$ and fixed depletant-to-lobe size ratio $q=\sigma_d/\sigma_A=0.12$. The depletant reservoir packing fraction mainly controls the strength of the attractive interaction while the  depletant size the range of the potential.}
\label{fig:potential}
\end{figure}

The colloidal tetrahedral particles as a whole interact via depletion interactions. To accurately  account for such interactions, each lobe in the tetrahedron is assigned a combination of a purely repulsive Mie potential $U_{Mie}(r)$ and an Asakura-Oosawa depletion potential $U_{AO}(r)$,~\cite{asakura1958interaction} which allows for a continuous treatment of the forces in molecular dynamics simulations. The total pair potential $U_{tot}(r)$  reads:
\begin{equation}
    U_{tot}(r)=U_{Mie}(r)+U_{AO}(r), 
    \label{eq:potential}
\end{equation}
with
\begin{eqnarray}
    \beta U_{Mie}(r)&& \nonumber \\ 
    &&\hspace{-1cm}=\begin{cases} 
    \frac{{4\epsilon}}{k_BT}\left[\left(\frac{\sigma_A}{r}\right)^{96}-\left(\frac{\sigma_A}{r}\right)^{48}\right]+\frac{\epsilon}{k_BT} & r \leq 2^{1/48}\sigma_A\\
    0 & r > 2^{1/48}\sigma_A,
    \end{cases}
\end{eqnarray}
where $\epsilon$ is set to $1~k_BT$ throughout this work,  and with 
%\begin{widetext}
%\footnotesize
\begin{eqnarray}
    \beta U_{AO}(r)& &\nonumber \\
    &&\hspace{-1cm}=\begin{cases} 
    -\eta_d^r\frac{\left(1+q\right)^3}{q^3} \\
    \times \left[1-\frac{3r}{2\sigma_A(1+q)}+\left(\frac{r^3}{2\sigma_A^3(1+q)^3}\right)\right] & r\leq\sigma_A+\sigma_d \\
    0 & r>\sigma_A+\sigma_d, 
    \end{cases}
\end{eqnarray}
\normalsize
%\end{widetext}
where $q=\sigma_d/\sigma_A$ is the size ratio between the depletants  of size $\sigma_d$ and the colloidal lobe particles of diameter $\sigma_A$,
and $\eta_{d}^r$ is the depletant reservoir packing fraction corresponding to a depletant  activity $z_d$ via $\eta_d^r=z_d\pi \sigma_d^3/6$. Finally, $\beta=1/k_BT$ denotes the inverse temperature with $k_B$ the Boltzmann constant. The parameters $q$ and $\eta_{d}^r$ control the characteristics of the potential, with $q$ mainly modulating the range of the potential and $\eta_{d}^r$ its depth as shown in Fig.~\ref{fig:potential}. 

In experiments, the central core monomer is coated with DNA which induces a strong short-range attractive interaction via the exposed area of the particle. Such interactions were found to promote self-assembly.
Here, we mimic this directional interaction $U_{DNA}(r)$ by allowing the central particles to interact through a ${96-48}$ Mie potential in which $\epsilon_{DNA}=10~k_BT$,
\begin{equation}
    \beta U_{DNA}(r)=\begin{cases} 
    \frac{{4\epsilon_{DNA}}}{k_BT}\left[\left(\frac{\sigma_B}{r}\right)^{96}-\left(\frac{\sigma_B}{r}\right)^{48}\right] & r \leq r_{cut}\\
    0 &  r > r_{cut}.
    \end{cases}
\end{equation}
Additionally, the potential is cut and shifted at $r_{cut}=1.15\sigma_B$. As the presence of DNA suppresses depletion, the particles in the center interact only with a Mie potential.~\cite{he2020colloidal} To assess the effect of DNA in the colloidal diamond assembly, we also run simulations in which the central particles interact via depletion only. 
In all cases,  the  tetrahedral colloid is treated as a rigid body. 

In order to explore the phase behavior of such colloidal molecules, we perform Langevin simulations in the $NVT$  ensemble using  LAMMPS,~\cite{thompson2022lammps} where $N$ denotes the number of particles, $V$ the volume of the system, and $T$ the temperature. We simulate $N=1080$ tetrahedral particles and we fix the side of the tetrahedron $a=0.7\sigma_A$.
We start our simulations from three to four different random low-density configurations which are slowly thermalized to $T^*=k_BT/\epsilon=1.0$ and progressively compressed to the desired final colloid density $\rho\sigma_A^3$, or packing fraction $\eta_c=\rho V_0$, with $\rho$ the number density and $V_0$ the volume of a single colloidal molecule, determined by accounting for the overlaps of the constituting five spheres. This is followed by an equilibration run for at least $t=1.0\times10^6 \delta t$, where $\delta t$ is the time step corresponding to $\delta t=0.001 \tau$ with $\tau=\sqrt{m \sigma ^2 / \epsilon}$ our time unit. Finally, a production run is made, with the system evolving for at least $2.5\times10^7 \delta t$. All simulations are performed in a cubic box with periodic boundary conditions. 

\subsection{Structural Characterization}
\label{subsec:structural}
In order to follow the nucleation and crystallization of cubic diamond, we need a criterion to distinguish between diamond and fluid-like particles. To this end, we first compute the correlation parameter $c_l(i,j)$ between the central particle of two neighboring colloidal molecules $i$ and $j$~\cite{neophytou2021self}
\begin{equation}\label{eq:corrparam}
\begin{split}
c_l(i,j) &=\frac{q_l(i) \cdot q_l^*(j)}{|q_l(i)||q_l(j)|} \\
&\hspace{-1cm}=\frac{\sum_{m=-l}^{l}q_{lm}(i)\cdot q_{lm}^*(j)}{\left(\sum_{m=-l}^{l}q_{lm}(i)\cdot q_{lm}^*(i)\right)^{1/2}\left(\sum_{m=-l}^{l}q_{lm}(j)\cdot q_{lm}^*(j)\right)^{1/2}} ,   
\end{split}
\end{equation}
where $*$ indicates the complex conjugate and $q_{lm}$ are the bond-orientational order parameters introduced by Steinhardt {\em et al.}~\cite{steinhardt1983bond}
\begin{equation}
    q_{lm}(i)=\frac{1}{N_b(i)}\sum_j^{N_b(i)}Y_{lm}(\mathbf{r}_{ij}),
\end{equation}
with $N_b(i)$ corresponding to the number of neighbors of particle $i$ and $Y_{lm}(\mathbf{r}_{ij})$ the spherical harmonics with total angular momentum $l$ and its projection $-l \leq m \leq l$.

The correlation parameter $c_l(i,j)$ takes characteristic values depending on the local arrangement of neighbouring molecules.  In the case of cubic diamond, its intrinsic symmetry corresponds to a tetrahedral one and hence, the expected number of neighbors per particle is four. Moreover, all the four bonds must have a staggered conformation. Here, we identify a staggered bond when $-1<c_3(i,j)<-0.8$.~\cite{romano2011crystallization, nguyen2015identification} In order to have a clear classification of the particles that are in a cubic diamond environment, we first calculate the nearest neighbors of each tetrahedral particle through a Voronoi construction, then we identify the four nearest neighbors and finally we calculate the corresponding $c_3(i,j)$. We remark that the four monomers at the vertices of each tetrahedral molecule $i$ are not accounted for in the calculation of $c_3(i,j)$, as they only serve to favor the assembly in a staggered conformation. If particle $i$ has four staggered bonds, it is considered part of a cubic diamond. We note that  when the cubic diamond nucleates into a grain or a gel-like structure particles that belong to the interface between the crystal and  fluid phase, tend not to fulfill this requirement. In order to include  them, we also consider particles that have three staggered neighbouring particles and at least one neighbor that is classified as cubic diamond to be part of the crystalline diamond structure.

\section{Results and discussion}
\label{sec:results}

\begin{figure*}
\includegraphics[width=1\linewidth]{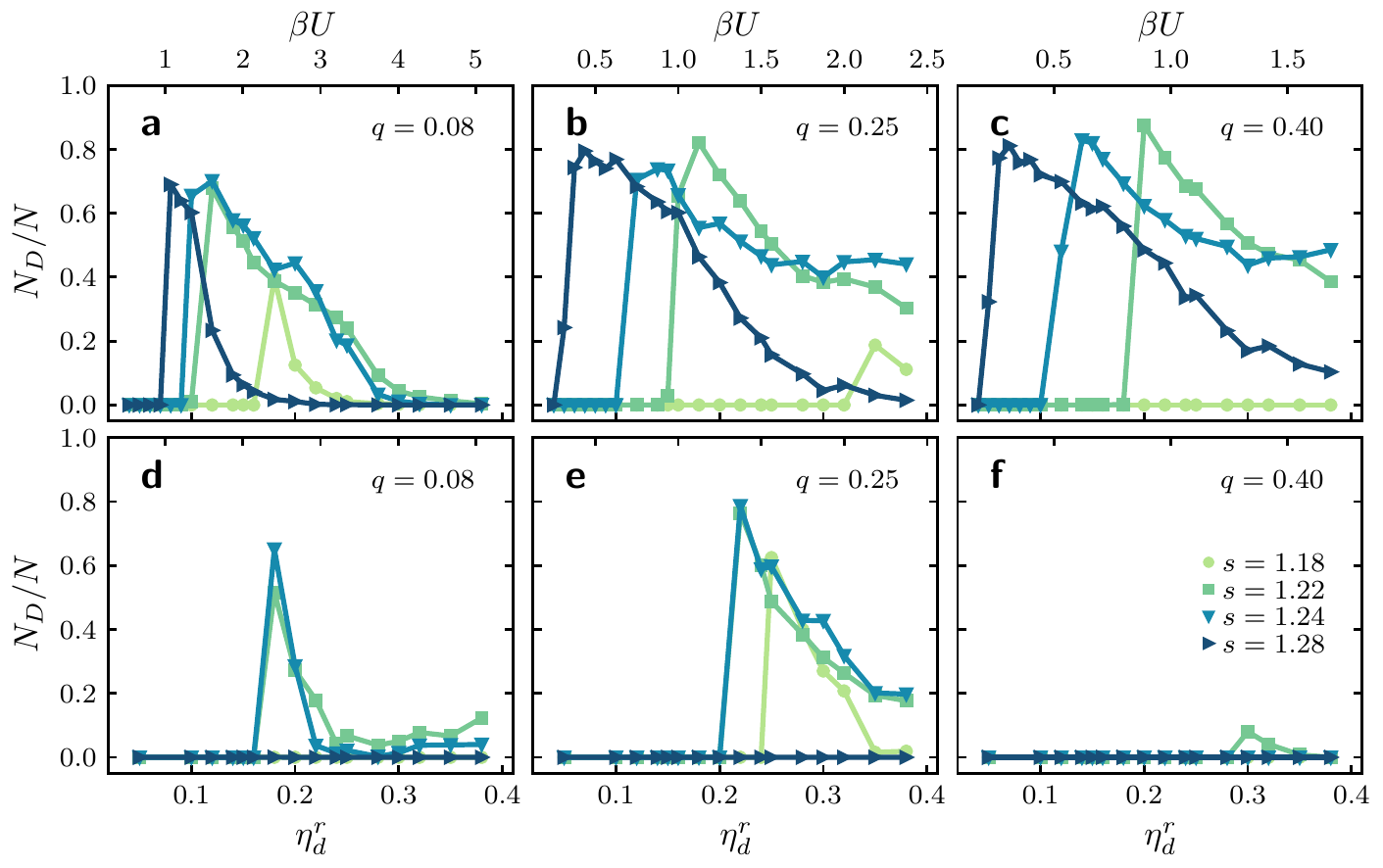}
\caption{Number fraction of particles $N_D/N$ identified as cubic diamond as a function of depletant reservoir packing fraction $\eta_d^r$ (lower axis) or interaction strength $\beta U$ (upper axis) for a system consisting of tetrahedral colloidal molecules, whose central core interaction (a-c) mimics the presence of DNA functionalization, or (d-f) presents solely depletion interactions, at  colloid number density $\rho\sigma_A^3=0.06$. Data are shown for different depletant-to-lobe size ratios $q=\sigma_d/\sigma_A=0.08, 0.25$ and $0.40$ (from left to right) and for core-to-lobe size ratios $s=\sigma_B/\sigma_A=1.18, 1.22, 1.24$ and $1.28$ as labeled.
}
\label{fig:cluster_s_q}
\end{figure*}

\subsection{The effect of both depletion and DNA interactions on the nucleation of cubic diamond}

We start our discussion by exploring the effect of tuning the core-to-lobe size ratio $s=\sigma_B/\sigma_A$ and the depletion interaction on the nucleation of cubic colloidal diamonds. In particular we focus on the effect of depletant-to-lobe size ratio $q=\sigma_d/\sigma_A$ and the depletant reservoir packing fraction $\eta_d^r$. We run $NVT$ simulations at fixed  colloid number density $\rho\sigma_A^3=0.06$, varying $s\in\{1.18,1.22,1.24,1.28\}$, $q\in\{0.08,0.25,0.40\}$ and $\eta_d^r \in \lbrack 0.05,0.40\rbrack$. 

For each simulation at different conditions we compute the number fraction of particles in a cubic diamond environment $N_D/N$, as described in Section \ref{subsec:structural}. In Fig.~\ref{fig:cluster_s_q}a-c we show $N_D/N$  for DNA-functionalized colloidal molecules as a function of $\eta_d^r$ for selected values of depletant-to-lobe size ratios $q$ and core-to-lobe size ratios $s$.
%The fraction of particles in a cubic diamond environment $N_D/N$ is reported in Fig.~\ref{fig:cluster_s_q}(a-c) as a function of $\eta_d^r$ for selected $q$ and $s$. 
For each $q$, we also report on the upper axes the corresponding interaction strength $\beta U$. 
%as the strength of the attractive interaction is an important parameter in controlling the assembly of the particles. 
Here, we take the interaction strength $\beta U=|\beta U_{min}(r)|$ with $\beta U_{min}(r)$ corresponding to the minimum of the total pair potential in Eq.~\ref{eq:potential} (see Fig.~\ref{fig:potential}). For the present system, where particles interact via  soft interaction  potentials, this appears to provide a more reasonable estimate than the contact energy as reported for instance in Ref.~\cite{fortini2008crystallization}.
In general terms, we first notice that by increasing $q$ the range of $\eta_d^r$, in which diamond-like structures are found, progressively widens. This effect is given both by the amount of diamond that is found for each $s$, and by the fact that at high $q$ large fractions of staggered arrangements are found for increasing values of depletant concentrations $\eta_d^r$. Overall, while for $q=0.08$ cubic diamond arrangements are mainly found for $0.05 < \eta_d^r < 0.25$, for $q=0.40$ they are found for  almost the entire $\eta_d^r$ range explored from $0.05$ to $0.40$. Therefore, an increase in the range of the potential, provided by larger $q$, effectively favors the assembly in staggered conformations, at all depletant concentrations. 

In spite of this, not all the sizes of the central core contribute equally to the formation of this type of structures, suggesting that a fine-tuning in the value of the core-to-lobe size ratio $s$ is actually crucial. In particular, we observe high fractions of particles arranged in a staggered conformation for very specific $s$, namely $s=1.22$ and $1.24$. In these cases, at appropriate conditions, even more than $80\%$ of the particles adopt a cubic diamond conformation. On the contrary, for a small size of the central core, less particles are found in diamond conformations, as reflected by  smaller values of $N_D/N$ and with the emergence of random structures. This is particularly evident for depletant-to-lobe size ratios $q \gtrsim 0.25$. This behavior can be explained by observing the structure of the particle as the size of the central core is varied, as depicted in Fig.~\ref{fig:model}. In fact, for small $s$, the directionality of DNA interactions is enhanced by the presence of four well separated triangular patches in between the lobes of the colloidal molecules. This implies that the nucleation of cubic diamond structures is achieved only for narrow $\eta_d^r$ ranges. On the other hand, when the size of the core particles is too large, the DNA interactions acquire a stronger isotropic character which is reflected in the loss of ordered staggered arrangements between colloidal molecules. 
This can be observed for the $s=1.28$ tetrahedral colloidal clusters, for which the range of $\eta_d^r$ where $N_D/N$ is non-zero is always smaller than the other two intermediate core-to-lobe size ratios $s$  analyzed.
In the latter case, the highest number of diamond arrangements is found for the lowest $\eta_d^r$ compared to the other size ratios, indicating that the nucleation is led by the strong DNA-mediated attractive interaction, and hence for large $s$ a smaller depletion attraction is required to achieve an ordered structure. We remind that all calculations presented in Fig.~\ref{fig:cluster_s_q} are performed for the same size of the tetrahedron side $a=0.7\sigma_A$. By changing this parameter, we simply expect a shift in the values of $q$, $\eta_d^r$ and $s$, although the observed phenomenology would be the same.

\subsection{The effect of solely depletion interactions on the nucleation of cubic diamond}
This picture highlights the non-trivial interplay between the interactions at stake: only for specific combinations of the parameters high yields of cubic diamond  can be obtained. To deepen our understanding on the role played by the DNA in the formation of cubic diamond structures, we analyze simulation runs for the same set of parameters but now for colloidal molecules whose core particles  only  present attractive depletion interactions as for the monomers in the lobes. The number fraction of colloidal diamond $N_D/N$ found for such systems is reported in Fig.~\ref{fig:cluster_s_q}d-f. By comparing with the previously described case, the overall amount of particles in a cubic diamond arrangement globally reduces and basically zeroes for $q=0.4$. However, for $q=0.25$, high fractions are still found for a wide range of $\eta_d^r$ and for a smaller range for $q=0.08$. The fact that under  such conditions we still find particles arranged as  colloidal diamond demonstrates that the shape of the particles in the presence of an attraction such as that given by the depletants suffices to create staggered conformations. From an experimental point of view this is certainly an advantage, since often the DNA functionalization may not be effective, besides typically involving a highly elaborated procedure. On the other hand, based on our evidence, it appears that the presence of DNA on the core particles gives a substantial contribution to generate colloidal diamond conformations with high yields and for a wide parameter space.

\begin{figure}[!ht]
\includegraphics[width=0.9\linewidth]{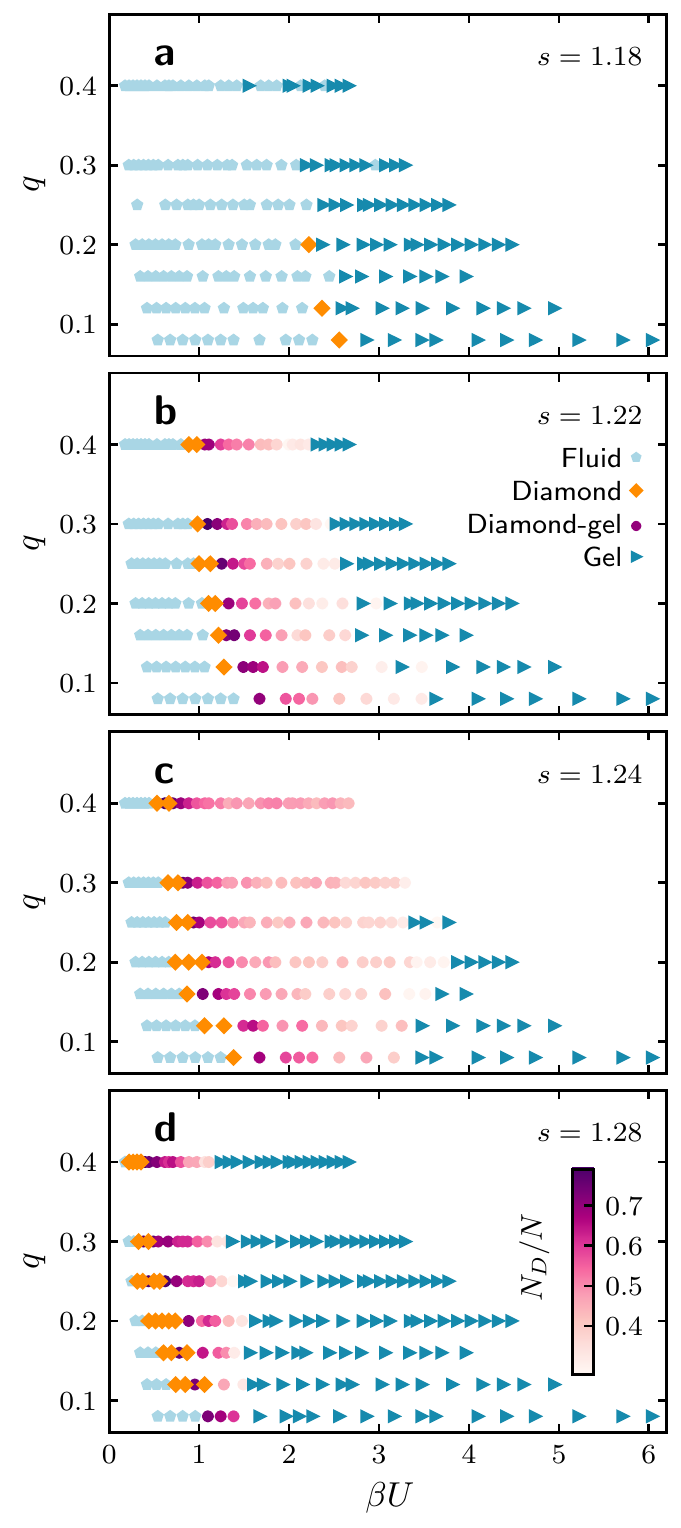} 
\caption{Kinetic phase diagram of  a system of DNA-functionalized tetrahedral  colloidal molecules at a colloid number density $\rho\sigma_A^3=0.06$ as a function of  depletant size $q=\sigma_d/\sigma_A$ and  interaction strength $\beta U$   for varying core-to-lobe size ratios $s=\sigma_B/\sigma_A$: (a) $1.18$, (b) $1.22$, (c) $1.24$ and (d) $1.28$. Four different phases are identified, namely cubic diamond crystals (orange diamonds), percolating cubic diamond gel-like structures (pink to purple shaded circles), disordered percolating gel networks (blue triangles) and an isotropic fluid phase (light blue pentagons). Different shades of pink denote different number fractions of cubic diamond $N_D/N$ in the percolating structures following  the color bar.
}
\label{fig:pd_q_eta}
\end{figure}

%%% Discussion about phase diagram DNA with all q and s

\subsection{Kinetic phase diagram as a function of depletant size and interaction strength}
In the case of colloidal particles that interact via  short-ranged attractive interactions, percolating gel networks can be formed due to the arrest of a  gas-liquid phase separation that is metastable with respect to a broad gas-solid transition.~\cite{lu2008eld,fortini2008crystallization,soto2022clusters} This non-equilibrium phenomenon of gelation has been studied in detail for a mixture of colloidal spheres and non-adsorbing polymer both in simulations and experiments.~\cite{lu2008eld,fortini2008crystallization,soto2022clusters} The arrest of this spinodal network is either due to crystallization at sufficiently low interaction strength $\beta U$, or caused by an attractive glass transition that meets the liquid binodal at high $\beta U$.~\cite{lu2008eld,fortini2008crystallization,soto2022clusters}

In our case, the use of large values of $\eta_d^r$ induces strong short-range attractive depletion interactions, and hence, we may also expect  under certain conditions the formation of not only compact crystal grains of cubic diamond, but also more branched and open structures, with the latter possibly retaining a certain degree of crystalline order. We therefore characterize in detail the structures that are formed for all the explored values of $q$ and $\eta_d^r$. Here and in the following we focus on the assembly of colloidal molecules whose core particles are functionalized with  DNA. 

\begin{figure*}
\includegraphics[width=1\linewidth]{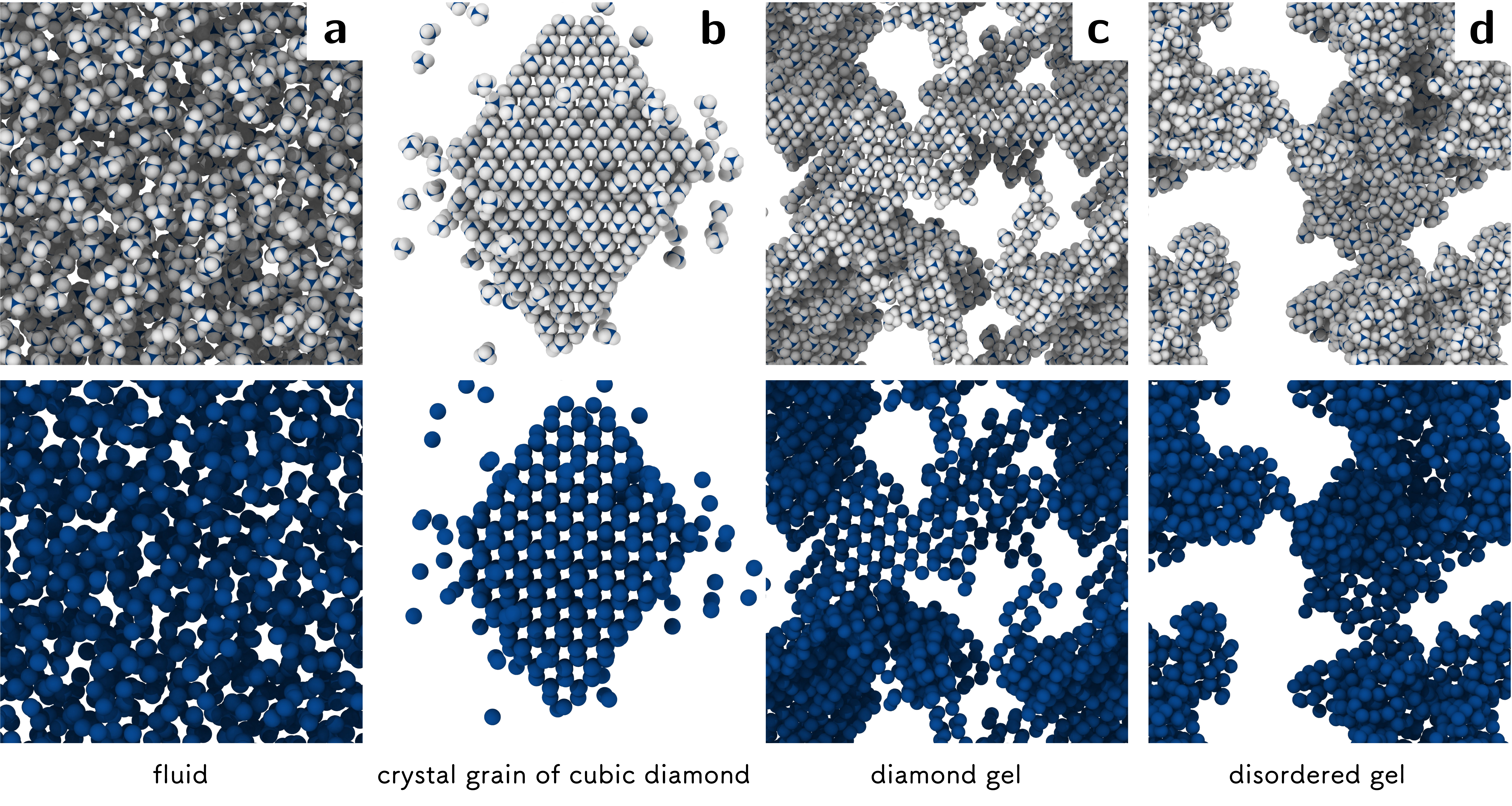}
\caption{Typical configurations obtained from simulations for the phases found for the tetrahedral colloidal molecules at a colloid number density $\rho \sigma_A^3=0.06$ for a core-to-lobe size ratio $s=\sigma_B/\sigma_A=1.24$ and depletant-to-lobe size ratio $q=\sigma_d/\sigma_A=0.12$: (a) isotropic fluid  phase, (b) crystal grain of cubic diamond with an octahedral shape, (c) percolating cubic diamond gel structure,  and (d) a disordered percolating gel network. Top panels report the full colloidal molecules, while bottom panels display only the core particles.
}
\label{fig:snapsphases}
\end{figure*}

To investigate whether or not gelation occurs in our system of tetrahedral colloidal molecules, we  first check under which conditions the system percolates. To this end, we first obtain the largest cluster formed by the colloidal molecules and check whether it spans the dimensions of the simulation box. Here, we consider two colloidal molecules to belong to the same cluster when the distance between their centers is less than a typical cut-off radius $|\mathbf{r}_i-\mathbf{r}_j|<r_c$, where we choose  $r_c\approx 1.6\sigma_A$, which allows for a correct identification of the percolating cluster. After having identified the extension of the largest cluster in each direction $l_i$, with $i=x,y,z$, we determine a system to be in a gel-like percolating network if $l_i>L-2\sigma_B$, where $L$ is the size of the box in one direction. We notice that even if the system forms a gel-like percolating network the amount of crystalline structures is still considerable. We  classify a system as a diamond gel-like percolating network when the system percolates and the fraction of particles labeled as diamond is $N_D/N \geq 0.25$, while for a simple gel the latter condition is not required. Finally, a system is considered as cubic diamond when $N_D/N \geq 0.25$ and it forms a non-percolating compact crystal grain.

Our results are summarized in Fig.~\ref{fig:pd_q_eta} in a kinetic phase diagram for a system consisting of DNA-functionalized tetrahedral colloidal molecules at a colloid number density $\rho\sigma_A^3=0.06$ as a function of the depletant-to-lobe size ratio $q=\sigma_d/\sigma_A$ and interaction strength $\beta U$ for various  core-to-lobe size ratios $s=\sigma_B/\sigma_A$.
For all core-to-lobe size ratios $s$ and depletant sizes $q$,  the system typically remains in a fluid phase at small interaction strengths $\beta U$, thus not nucleating any crystal phase. Depending on the  size ratio $s$, the fluid region can be found for a wide range of interaction strengths $\beta U$ and depletant size ratios $q$, as for instance for $s=1.18$ where the phase diagram is dominated by an isotropic fluid phase. A typical configuration of the fluid of tetrahedral colloidal molecules is reported in Fig.~\ref{fig:snapsphases}a. Upon increasing $\beta U$, the fluid is followed by the nucleation of cubic diamonds. The set of parameters where nucleation occurs into a well-defined diamond crystal grain without percolation is relatively narrow, corresponding to an interaction strength $0.5 \leq \beta U \leq 2.5$ depending on $s$. In general, at larger interaction ranges, that is for larger depletant sizes $q$, the concentration of depletants  needed for nucleating the cubic diamond is lower, as less cohesive energy is needed to form four staggered bonds per particle. This is in contrast to smaller values of $q$, for which slightly stronger interactions are needed to find good nucleation. As a consequence of the intrinsic tetrahedral shape of the colloidal molecules, we frequently find the formation of crystal grains with an octahedral shape as reported in Fig.~\ref{fig:snapsphases}b. For the case where all monomers interact via depletion only, the region where diamond grains are nucleated is further restricted (see Fig. S1), supporting the role of DNA to enhance crystallization.

Close to the region in the kinetic phase diagrams of Fig.~\ref{fig:pd_q_eta} where non-percolating cubic diamonds are formed,  we observe the formation of a percolating network at slightly higher interaction strength $\beta U$. This phenomenon has been previously observed for other systems presenting short-ranged attractive interactions and typically signals the presence of a metastable gas-liquid phase transition.~\cite{lu2008gelation, fortini2008crystallization}
An interesting aspect we find in this region concerns the fact that the structure of the percolating network is not disordered. On the contrary, most of the particles have  neighbors in a staggered conformation, effectively stabilizing a cubic diamond crystal in a mesoscopic  gel-like structure, as shown in Fig.~\ref{fig:snapsphases}c. This type of structure is found for a wide range of interaction strengths $\beta U$ for $s=1.22$ and $1.24$, and effectively constitutes the most dominant phase for these two size ratios $s$. This demonstrates the importance of the directionality of the central core particle in the formation of staggered structures. In fact, the moment this aspect is lacking, as in the case of the two size ratio limits, fully disordered conformations prevail. 
In all cases, as the interaction strength gets stronger, the arrest is more pronounced and the crystallinity is gradually lost, leading to a fully disordered gel network (see Fig.~\ref{fig:pd_q_eta}). In case the central particle is not coated with DNA, the interaction range at which disordered phases are found is lower compared to the case discussed here, and starts already at $\beta U \approx 2$ (see Fig. S1).

To summarize, the assembly of tetrahedral colloidal molecules into cubic diamond requires a precise choice in terms of interaction strengths, and hence in the concentration of solutes acting as depletants. When deviating from these values, arrested phases are found, at first maintaining an intrinsic crystalline structure before forming disordered phases at higher interaction strengths. A less stringent requirement seems to emerge instead in terms of the depletant size $q$, provided the core-to-lobe size ratio $s$ is chosen appropriately. We note that we find no other crystalline structures for the parameters and ranges that we explored in our simulations.

\subsection{Structural features of the different phases}

\begin{figure}[t!]
\includegraphics[width=1\linewidth]{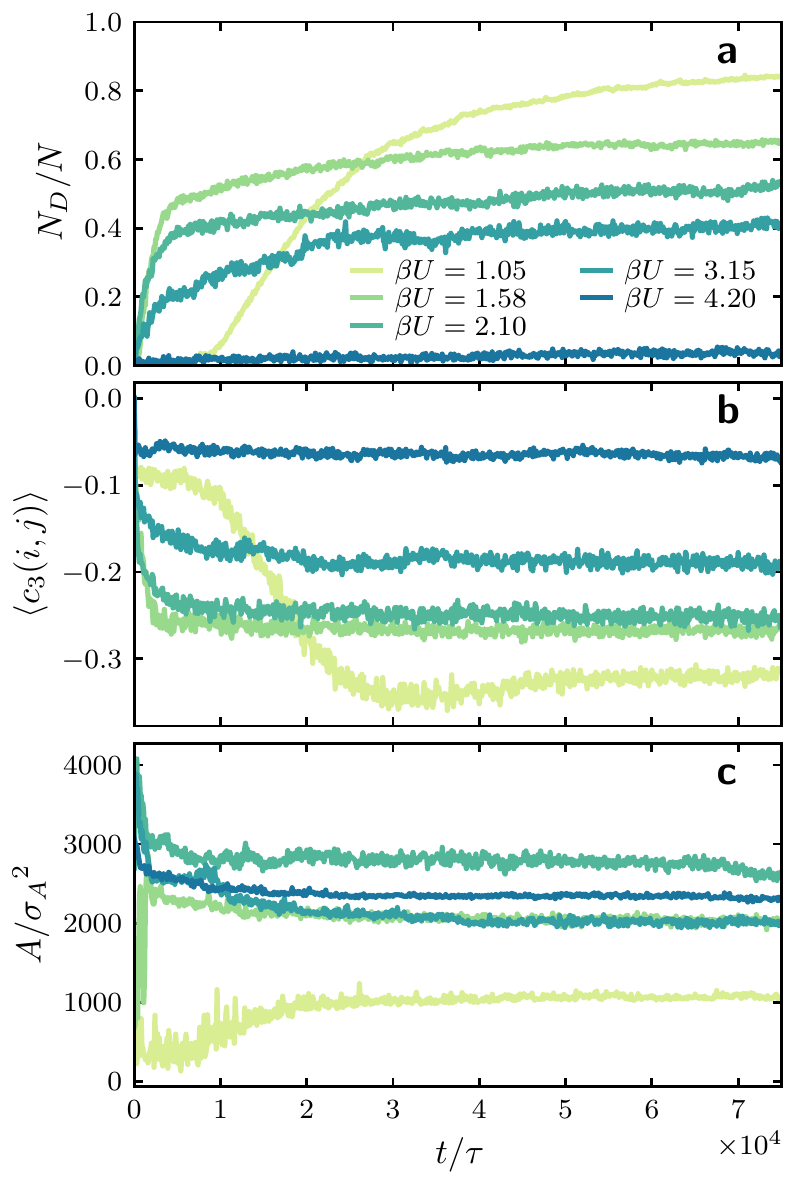} 
\caption{(a) Number fraction of particles $N_D/N$ identified as cubic diamond, (b) average correlation parameter $\left< c_3(i,j) \right>$ and (c) surface area $A/\sigma_A^2$ of the assembled structures calculated by means of a surface mesh, as a function of  simulation time $t/\tau$ for a system of DNA-functionalized tetrahedral colloidal molecules with a core-to-lobe size ratio $s=1.24$, depletant-to-lobe size ratio $q=0.12$, and at colloid density $\rho \sigma_A^3=0.06$ for varying  interaction strengths $\beta U$ as labeled.
}
\label{fig:solid}
\end{figure}

In the previous section, we observed how the kinetic phase diagrams display both  crystalline diamond structures and  arrested states as a result of the strong interactions caused by the depletant agents. 
We now explore in detail such phases by analyzing their peculiarities and the differences in their local structure.

\begin{figure*}
\includegraphics[width=1\linewidth]{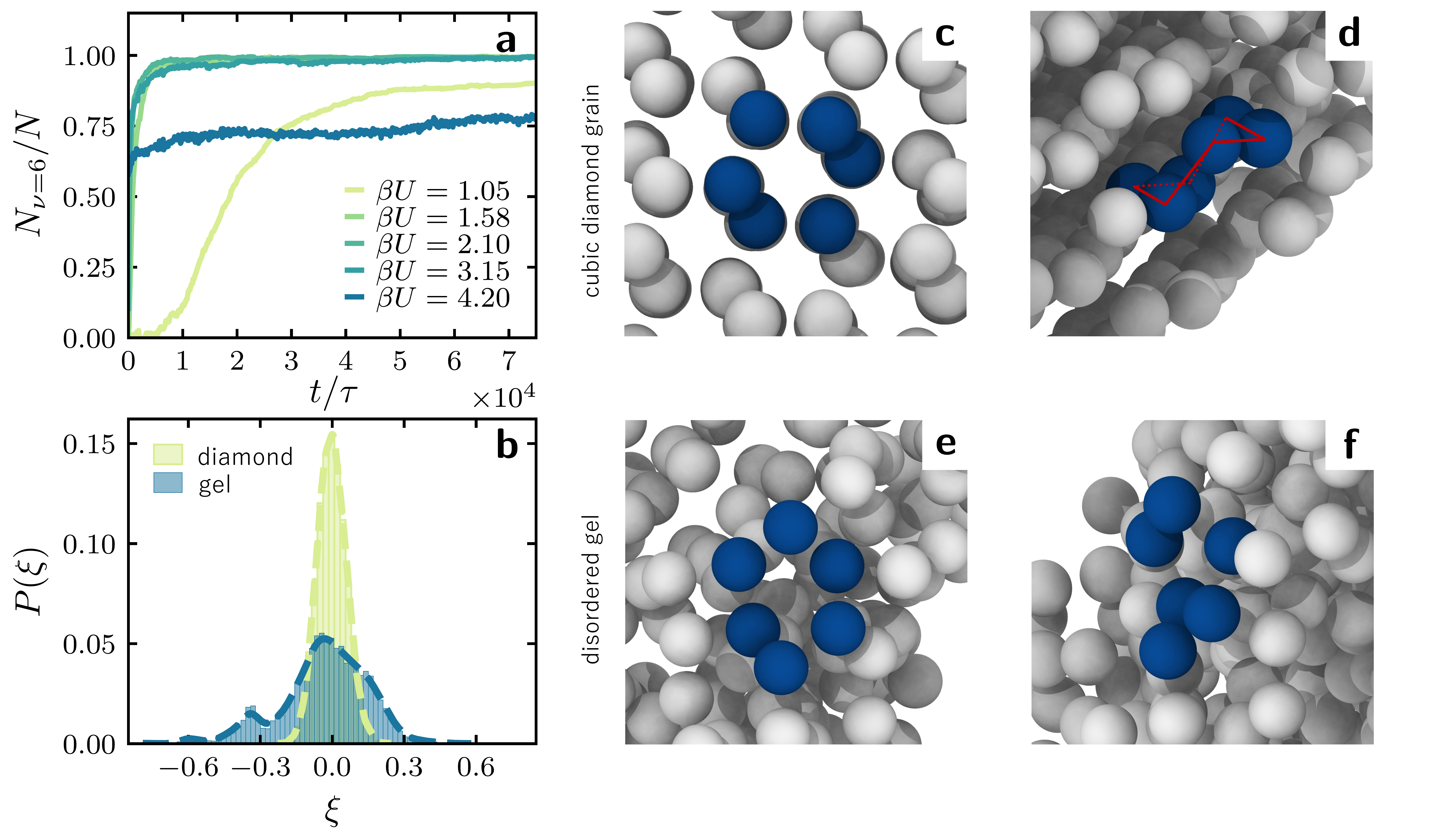} 
\caption{Structural characterization of the diamond crystal and the percolating gel structures. (a) Fraction of six-membered rings $N_{\nu=6}/N$ as a function of simulation time $t/\tau$ for a system consisting of tetrahedral colloidal molecules with a core-to-lobe size ratio $s=1.24$, depletant-to-lobe size ratio $q=0.12$, at a colloid number density  $\rho \sigma_A^3=0.06$ for varying interaction strength $\beta U$ as labeled. (b) Probability distribution of the angular strain $P(\xi)$ of rings consisting of  six members for a crystal grain of cubic diamond and for a disordered gel configuration. (c-d) Typical simulation configurations of a diamond structure in which two different perspectives of  six-membered rings are shown. Red lines highlight the chair-like conformation of the ring. (e-f) Typical simulation configurations of  six-membered rings belonging to a gel-like structure. The chair-like conformation and the corresponding internal angle distributions are no longer retained in this structure.
}
\label{fig:rings}
\end{figure*}

The cross-over from the region where diamond grains are nucleated to the region in which percolating systems are formed is narrow and strongly depends on the precise value of the depletant size $q$ and depletant reservoir packing fraction $\eta_d^r$. In order to shed light on the mechanism of nucleation of cubic diamond and the difference between percolating and non-percolating structures, we focus on a set of parameters $s$ and $q$ for which we find all the phases we are interested in and we characterize the structural evolution of the system.
%as a function of time. 
In particular, we analyze selected values of $\eta_d^r$ for a system with $s=1.24$ and $q=0.12$, for a fixed density of the colloidal molecules $\rho \sigma_A^3=0.06$ as for the case reported in Fig.~\ref{fig:pd_q_eta}c. 

We inspect the number fraction of particles in a diamond environment $N_D/N$ as a function of time and report the results  in Fig.~\ref{fig:solid}a.
%We start our analysis by inspecting the fraction of particles in a diamond environment $N_D/N$ as a function of time for different conditions, we show this results in Fig.~\ref{fig:solid}a. 
It is immediately evident that  we can identify three different regimes, for low, intermediate and high $\beta U$. The nucleation of cubic diamond grains corresponds to the first regime, where the fraction of molecules in a staggered conformation slowly increases. This process allows the particles to slowly rearrange by progressively forming a well-ordered compact structure (see Video S1). As the interaction strength $\beta U$ increases, the growth in $N_D/N$ is increasingly pronounced and reaches sooner a plateau. 
%the assembly process gets faster and $N_D/N$ rapidly reaches a plateau. 
In this regime the interactions are such that locally an ordered crystal can still be obtained, while being nucleated into a branched gel-like structure. 
The concurrent presence of disordered arrangements is consistent with an overall lower fraction of diamond at long times. At even higher depletant concentrations,  for $\beta U \gtrsim 4$, the system is immediately arrested in a disordered conformation and no diamond-like structure is nucleated over time. 

As expected, this picture is consistent with the representation given by the local bond order parameter for cubic diamond. In particular, in Fig.~\ref{fig:solid}b, we plot the average correlation parameter $\left< c_3(i,j) \right>$ as a function of time, where the average is performed over $N_b=4$  neighbors. 
For increasing interaction strengths, $\left< c_3(i,j) \right>$ tends to values closer to zero, reflecting the fact that fewer and fewer particles are in a cubic environment, as $c_3(i,j)$ progressively deviates from $-1$. However, it is important to note that the mean value of the correlation function, as calculated here, is also affected by the presence of particles at the interface. As a result,  $\left< c_3(i,j) \right>$ also shifts towards values closer to zero due to a lower number of first neighbors in a staggered conformation. To understand whether the increase is only due to a loss of crystalline environment and the role played by the interface, we calculate for each state point the surface area of the structure being formed over time by means of a surface mesh~\cite{stukowski2014computational} that encompasses the largest cluster. As  can be observed in Fig.~\ref{fig:solid}c, the formation of a crystalline grain is associated to a smaller surface area given that the colloidal molecules are arranged in a more compact structure. For larger interaction strengths, larger and comparable surface areas are retrieved, consistent with the presence of more expanded structures. Thus, except for the lowest interaction strength where the compact diamond crystals are stabilized, the increase in the value of the correlation parameter $\left< c_3(i,j) \right>$ for $\beta U \gtrsim 1.5$ can essentially be attributed  to a progressive loss of crystallinity in the structure rather than to a larger amount of particles at the surface.

The nucleation and formation of  crystalline and arrested phases is often characterized by the emergence of characteristic local structures.~\cite{coli2021crystal, griffiths2017local, marin2020tetrahedrality}
%Crystal nucleation typically involves the formation of recurrent patterns within the structure. 
For instance, the formation of ring motifs between neighboring particles was identified in the nucleation pathways of patchy particles into open crystal structures.~\cite{rao2020leveraging, neophytou2021self} In particular, for cubic diamond and its polymorphs, the inner structure presents rings consisting of $6$ or $8$ members.~\cite{neophytou2021self} 
Here, we explore how the formation of such structures occurs in our systems of anisotropic tetrahedral colloidal clusters. In particular, we focus exclusively on six-membered rings formed by the central monomers of the tetrahedral colloidal molecules, which are known to retain a chair-like or staggered conformation in perfect diamond crystals. To this end, we first identify the nearest neighbours through a cutoff distance $r_c=1.6\sigma_A$, corresponding to the first minimum of the radial distribution function of a cubic diamond formed by our model, ensuring that at most four nearest neighbors are taken into account. We verified that a variation of $r_c$ in between the first and second peak of the radial distribution function does not affect the correct identification of the neighbors and thus of the rings (not shown). A six-membered ring is identified when, starting from a given particle, it is possible to return to the same one, passing through five other particles that are consecutive first neighbors to each other. To shed light on the difference between the structures  found at different depletion strengths, we compute the fraction of particles belonging to a six-membered ring $N_{\nu=6}/N$ as a function of time. In Fig.~\ref{fig:rings}a we show $N_{\nu=6}/N$ for different interaction strengths $\beta U$.
%The evolution of the fraction of rings with $\nu=6$, $N_{\nu=6}/N$, is reported in Figure~\ref{fig:rings}a. 
For low and intermediate interaction strengths, corresponding to low and intermediate depletant concentrations, the curves resemble closely  the trends of $N_D/N$ shown in Fig.~\ref{fig:solid}a. A progressive growth in $N_{\nu=6}/N$ is indeed found for the set of parameters that leads to crystalline grains, while  systems that are percolating display  a quick saturation at short times. For the highest interaction strength, instead, the number of six-membered rings is always non-zero and essentially constant over time. Since this type of rings occurs in all phases, it is therefore reasonable to expect their internal conformations to be very dissimilar. 
In particular, the internal angle found between particles in a perfect chair-like conformation amounts to $\theta_{ijk}=109.5^{\circ}$.~\cite{brown2014introduction} If the internal angle $\theta_{ijk}$ deviates from this value, the system acquires a so-called angular strain $\xi = \theta_{ijk}-109.5^{\circ}$.~\cite{wiberg1986concept} We  report in Fig.~\ref{fig:rings}b the angular strain probability distributions per particle for a crystalline cubic diamond ($s=1.24$, $q=0.12$, $\eta_d^r=0.12$) and for a disordered gel ($s=1.24$, $q=0.12$, $\eta_d^r=0.40$). Other probability distributions for intermediate $\beta U$ are reported in Fig. S2. As expected, the angular strain for diamond structures is distributed around zero, with most of the particles in the crystal grain retaining six-membered rings in a chair-like conformation. As the strength of the interaction increases, the rings tend to get distorted, thus introducing a larger angular strain. Consequently, the distribution broadens and once the network looses crystallinity it is characterized by the emergence of a second peak at around $-0.35$. Therefore, even if rings with six members are present for all phases, structures in an arrested phase retain a completely different conformation owing to the random arrangement of the building blocks. The differences between the six-membered rings for the two analyzed phases are clearly visible from the simulation snapshots reported in Fig.~\ref{fig:rings}c-d for a perfect diamond crystal grain and  in Fig.~\ref{fig:rings}e-f for a disordered gel-like  configuration.

\begin{figure}
\includegraphics[width=1\linewidth]{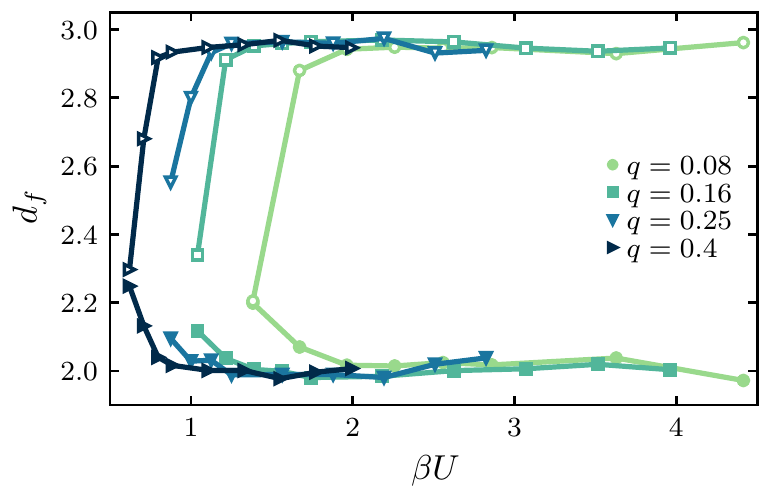} 
\caption{Fractal dimension $d_f$ as a function of the interaction strength $\beta U$ for different values of the depletant-to-lobe size ratio $q=\sigma_d/\sigma_A$ as labeled. Data are shown for a system of $N=10800$ particles where the colloidal molecules have a core-to-lobe size ratio $s=\sigma_B/\sigma_A=1.24$. Closed symbols relate to the fractal dimension $d_f$ at short length scales, while open symbols are for large length scales.}
\label{fig:fd}
\end{figure}

To further characterize the structural changes occurring across phases, we calculate the fractal dimension $d_f$ of the gel-like structures, making use of the box counting algorithm.~\cite{gagnepain1986fractal,ruiz2020tuning} To this aim, after having identified the largest cluster, we subdivide the simulation box into cells of varying size $\lambda$ and we count the number of cells $N_c$ that contains at least one colloidal molecule. The fractal dimension is then obtained from
\begin{equation}
\label{eq:fd}
\log N_c(\lambda) = - d_f \log \lambda.
\end{equation}
By fitting Eq.~\ref{eq:fd} in two different regimes, for large and small $\lambda$, it is possible to obtain information about the system at different length scales, corresponding to the internal structure and  the mesoscopic scale.~\cite{griffiths2017local} Here, we discriminate between the two regimes by taking a cross-over value $\lambda=8\sigma_A$. An example of the fitting procedure and a schematic representation of the box counting method for the two $\lambda$-regimes are reported in Fig. S3.
For a more reliable estimate of this quantity, the fitting is performed on simulations consisting of  $N=10800$ colloidal molecules.

The fractal dimension for the two different length scales is reported in Fig.~\ref{fig:fd} for a system of colloidal molecules with a core-to-lobe size ratio $s=1.24$ at different depletant-to-lobe size ratios  $q$ as a function of the  interaction strength  $\beta U$.
For the smallest interaction strengths analyzed,  $d_f$ takes similar values at large length scales,  always below $2.6$ for all $q$. Globally, this indicates a compact structure, which therefore does not extend throughout the box leaving large portions of empty space, as expected for a grain similar to the one shown in Fig.~\ref{fig:snapsphases}b. At the same time, for small length scales, $d_f$ takes the highest values at low interaction strengths. This indicates the tendency of  local structures to have a conformation with a large homogeneity of the system, owing to the still large amount of diamond crystal that is present. 
Moving to $\beta U > 2$, the formation of percolating and disordered structures prevails and the trend of the fractal dimension reflects this phenomenon. In fact, on large scales, $d_f$ tends to approach the highest value, suggesting that globally the structure occupies the whole box. This picture is fully consistent with that depicted in the previous sections for describing the branched and porous structure that comes along with the percolating gel network. At small length scales, the local disorder and the gradual decrease in the amount of crystal lower the fractal dimension and set it close to $2$.

\subsection{Kinetic phase diagram as a function of colloid density and interaction strength}

\begin{figure}[t]
\includegraphics[width=1\linewidth]{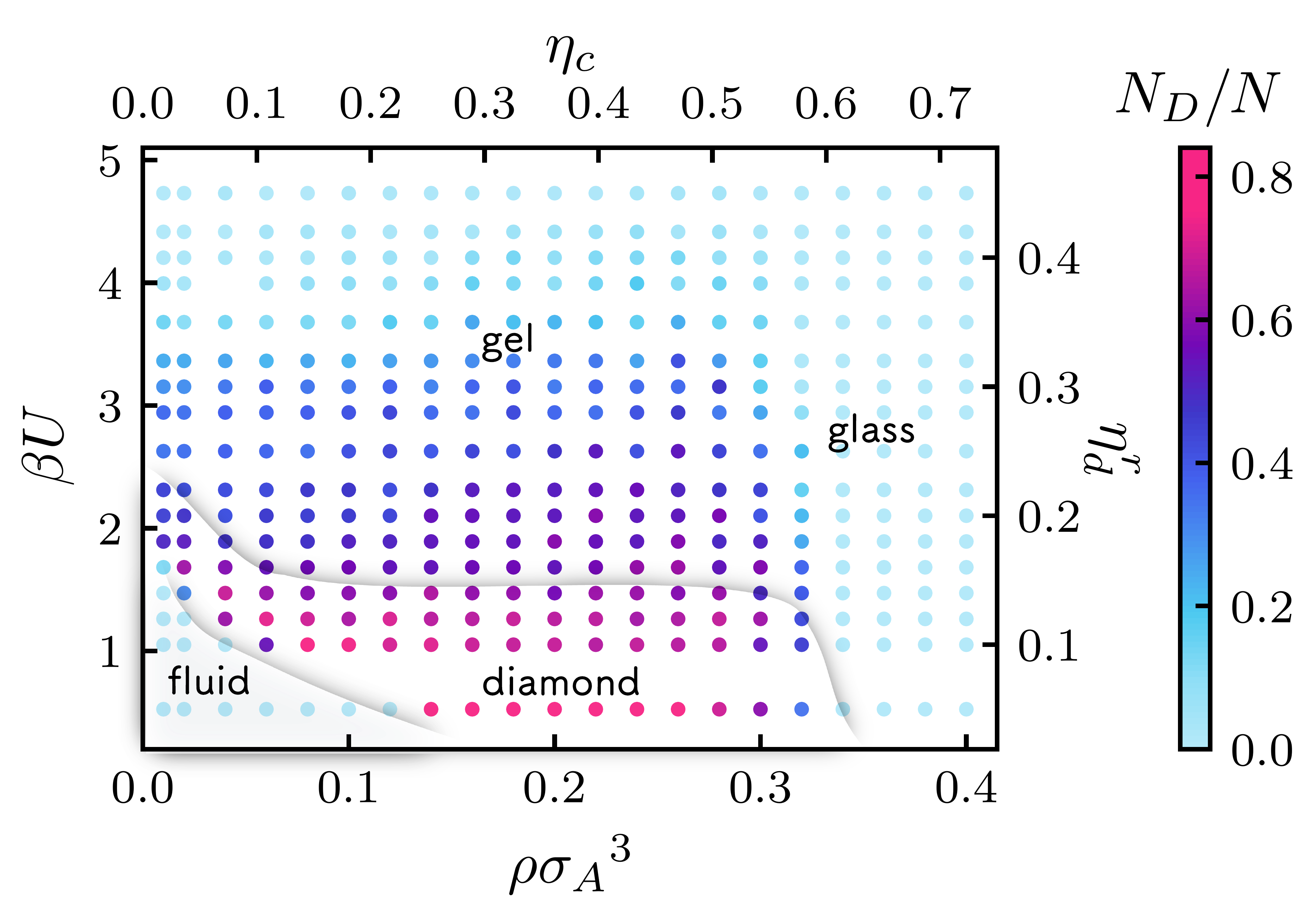} 
\caption{Kinetic phase diagram for 
a system of DNA-functionalized tetrahedral colloidal molecules with a core-to-lobe size ratio $s=\sigma_A/\sigma_B=1.24$ and a depletant-to-lobe size ratio $q=\sigma_d/\sigma_A=0.12$ as a function of interaction strength $\beta U$ (left axis) or depletant reservoir packing fraction $\eta_d^r$ (right axis), and colloid number density $\rho \sigma_A^3$ (lower axis),  or colloid packing fraction $\eta_c$ (upper axis). The color bar on the right indicates the fraction of particles in a diamond environment $N_D/N$.} 
\label{fig:PD}
\end{figure}

We now explore in detail the kinetic phase diagram for a system  of  tetrahedral colloidal molecules as a function of  colloid number density $\rho \sigma_A^3$ and interaction strength $\beta U$. To do so, we select a core-to-lobe size ratio $s$ and a depletant size $q$ for which we observe the formation of crystal  grains of cubic diamond at low densities. In particular, we run simulations for $s=1.24$ and $q=0.12$, and vary  the depletant reservoir packing fraction $\eta_d^r$ and the colloid density  $\rho\sigma_A^3$, respectively. With this approach we range from a dilute state to a highly packed system. We present the kinetic phase diagram for $q=0.12$ in Fig.~\ref{fig:PD} and for $q=0.25$ in Fig. S4. 
The color code in the phase diagram indicates the number fraction $N_D/N$ of particles in a cubic diamond environment, whose estimate is calculated based on the correlation parameter as before. 

The structures that are identified are consistent with those observed in Fig. \ref{fig:pd_q_eta} at low densities. Representative simulation configurations of high-density structures are displayed in Fig. S5. Formation of crystalline diamond is observed for a wide range of packing fractions of colloidal molecules, varying from the low density  $\rho\sigma_A^3 = 0.06$ analyzed in the previous Sections up to a colloid density of $\rho \sigma_A^3 \approx 0.35$, corresponding to a colloid packing fraction range of $0.05 \leq \eta_c \leq 0.6$, for weak depletion attractions. In comparison to simple tetrahedral patchy particles,~\cite{romano2010phase} the range in which colloidal diamond is found is almost double in terms of packing fractions. In this sense, a key role is played by the structure of the particles. While in the case of tetrahedral colloidal molecules  conformations not involving staggered bonds are actually disfavored, patchy particles could stabilize different crystal structures, such as body-centered cubic (bcc) and face-centered cubic (fcc) phases, at similar or higher packing fractions, respectively. For the tetrahedral colloidal molecules, such phases are not observed at any concentration. In terms of the strength of the interaction potential, this remains rather narrow. Only  for low colloid densities  crystalline grains of diamond are found  for relatively high interaction strengths $\beta U \approx 2.5$. At higher colloid densities fully crystalline diamond structures are only found for lower $\beta U$.

The central region of the phase diagram is  characterized by a lower  number fraction  $N_D/N$ of colloidal diamond. In this region the conformation of the crystalline diamond become more gel-like. For $\beta U \approx 1.8$, both ordered and disordered states coexist. By further increasing the interaction strength, a fully disordered structure emerges as expected. This change is also reflected in the diffraction pattern shown in Fig. S6, as compared to that of crystalline diamond. 
For high colloid densities $\rho \sigma_A^3$ and sufficiently high $\beta U$, the system is mostly in an arrested glassy state. Under these conditions, the depletion attractions are so strong that the system becomes arrested due to an attractive glass transition.  The transition to this state finds evidence both in the diffraction patterns and in the mean squared displacement reported in Fig. S6. Accordingly, the fraction of colloidal molecules in a diamond environment found in this region is essentially negligible. 
Finally, we note that by changing the size of the central particle or  the size of the depletants, the phase diagram does not present relevant changes at a qualitative level. Only some minor variations are observed in the range where  the different phases are found. 

\section{Conclusions}
\label{sec:conclusion}

The self-assembly of cubic diamond  is still an elusive process which typically involves the use of anisotropic particles and directional interactions. 
Here, we explored in detail the self-assembly behavior of a model of tetrahedral colloidal molecules based on a recent experimental realization that was  found to nucleate exclusively into cubic diamond.~\cite{he2020colloidal} 

For this purpose,  we employed in our simulations a realistic set of interactions,  including not only depletion interactions between the monomers at the vertices of the tetrahedron but also strong attractive interactions between the central core particles due to the presence of DNA. We found that the latter plays an important role in the formation of crystalline diamond. In particular, as compared to  building blocks where the core is DNA-functionalized, crystalline phases can be nucleated for a wider set of parameters, by modifying the characteristics of the depletants. In fact, by changing  the concentration or the  size of the depletants, the morphologies of the self-assembled structures can be tuned  from an isotropic fluid phase, via a colloidal crystal  to a gel-like structure. From an experimental point of view, one could envision that the characteristics of the different structures may be altered in situ by means of microgel particles made of a thermoresponsive polymer, that act as size-tunable depletants.~\cite{rossi2011cubic,rossi2015shape,xing2015depletion,bergman2018new}
%by means of a thermoresponsive polymer acting as a depletant, such as a nanogel particle~\cite{rossi2011cubic,rossi2015shape,xing2015depletion}. 

However, it should be emphasized that the formation of cubic diamond crystalline structures, which could find use in photonic crystals, remains limited to a relatively small range of  interaction strengths, which is further restricted as the colloid density increases.  Therefore, to avoid the formation of disordered gel structures and arrested states, it is essential to have a precise control on the interaction potentials and the thermodynamic state. We found indeed that a small deviation from the optimal conditions for nucleation and crystallization  leads, both at low and intermediate colloid densities, to the formation of percolating structures with  different degree of crystallinity. This is consistent with earlier observations for a system of colloidal spheres that interact with short-range depletion interactions, where a subtle interplay is found between crystallization, gelation, and vitrification.~\cite{lu2008gelation,fortini2008crystallization} The structural analysis we carried out for these structures, especially for $\beta U \gtrsim 3$, reveals a structure that percolates throughout the entire box. In addition,  the loss of crystallinity at high interaction strengths $\beta U$ is accompanied by an arrangement of colloidal molecules where the typical six-membered rings of the crystalline cubic diamond are distorted. At high $\beta U$,  the attractive interactions due to the presence of the DNA at the surface of the core particles are  no longer able to stabilize crystalline conformations and interactions occur mainly through the lobes.

In conclusion, this study elucidates under which  conditions the formation of cubic diamond from a system of tetrahedral clusters is favored. Our results may find use in experimental and application studies where the formation of structures with photonic properties is required. We hope that our results  inspires future experiments, where  the role of the depletion interactions is fully exploited in the assembly of cubic diamonds.

\bigskip

\begin{acknowledgments}
We thank David J. Pine for valuable discussions. The authors acknowledge funding from the European Research Council (ERC) under the European Union's Horizon 2020 research and innovation program (Grant agreement No. ERC-2019-ADG 884902, SoftML).
\end{acknowledgments}

\section*{Author Contributions}

S.M.-A. and F.C. contributed equally to this work.

S.M.-A.: Conceptualization, Methodology, Software,  Validation, Formal Analysis, Investigation, Data Curation, Visualization, Writing - Original Draft, Writing - Review and Editing. F.C.: Conceptualization, Methodology, Software, Validation, Formal Analysis, Investigation, Data Curation, Visualization, Writing - Original Draft, Writing - Review and Editing. M.D.: Conceptualization, Resources, Supervision, Project administration, Funding acquisition, Writing - Original Draft, Writing - Review and Editing.

\section*{Conflict of interest}
The authors have no conflicts to disclose.

\section*{Data Availability}
The data that support the findings of this study are available from the corresponding author upon reasonable request.

\clearpage
\newpage
%\widetext
%\begin{widetext}
%\begin{center}
\onecolumngrid

\large
%\noindent\textbf{Guiding the self-assembly of colloidal diamond\\ \bigskip Supporting Information}
\section*{\Large G\titlelowercase{uiding the self-assembly of colloidal diamond}\\ \smallskip S\titlelowercase{upporting Information}}
%\title{Guiding the self-assembly of colloidal diamond}
\normalsize

\noindent Susana Marin-Aguilar\textsuperscript{ 1}, Fabrizio Camerin\textsuperscript{ 1}, Marjolein Dijkstra\textsuperscript{ 1}\\
\medskip
\small
\textit{%
\textsuperscript{1}Soft Condensed Matter, Debye Institute for Nanomaterials Science, Utrecht University, Princetonplein 5, 3584 CC Utrecht, The Netherlands
}

\normalsize

\renewcommand{\thefigure}{S\arabic{figure}}\setcounter{figure}{0}

\section*{KINETIC PHASE DIAGRAM FOR TETRAHEDRAL COLLOIDAL MOLECULES WITHOUT DNA FUNCTIONALIZATION}

\begin{figure*}[b]
\includegraphics[width=0.85\linewidth]{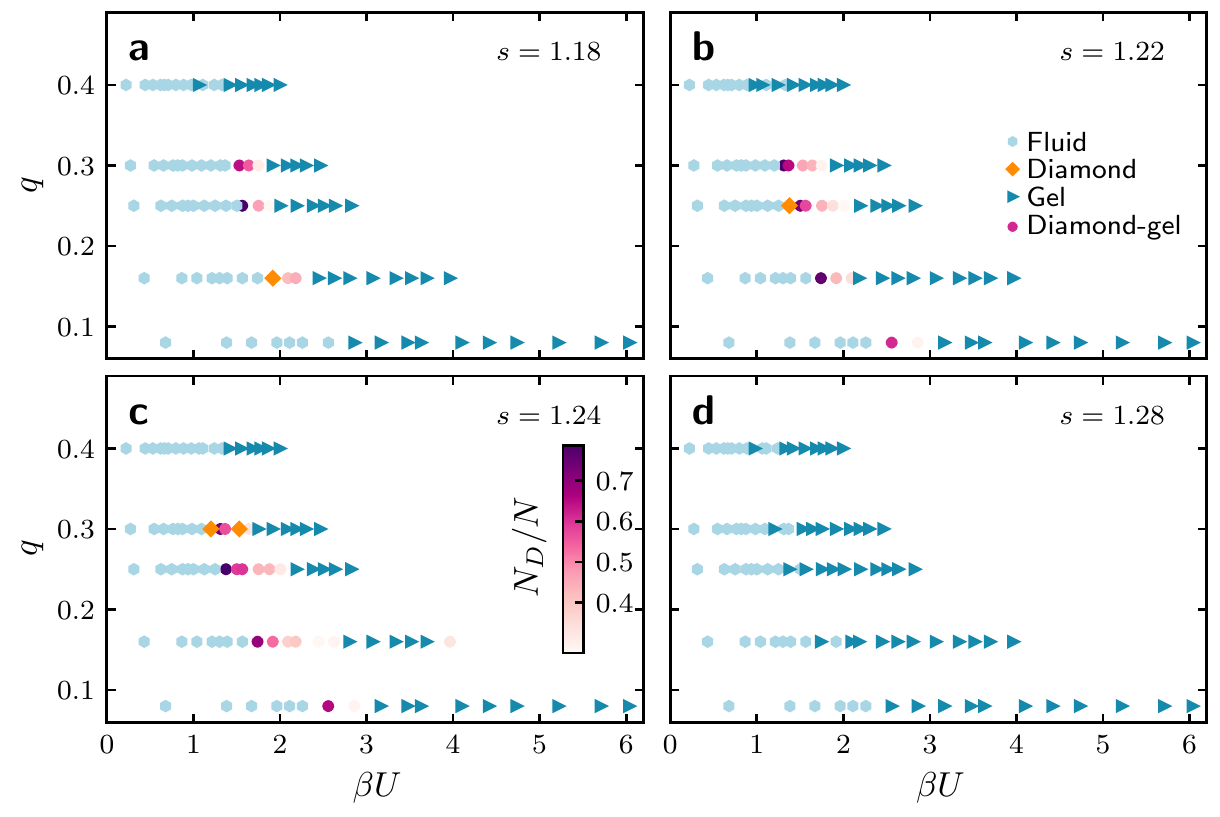}
\caption{Kinetic phase diagram of a system of tetrahedral  colloidal molecules without DNA functionalization at a colloid number density $\rho\sigma_A^3=0.06$ as a function of  depletant size $q=\sigma_d/\sigma_A$ and  interaction strength $\beta U$   for varying core-to-lobe size ratios $s=\sigma_B/\sigma_A$: (a) $1.18$, (b) $1.22$, (c) $1.24$ and (d) $1.28$. Four different phases are identified, namely cubic diamond crystals (orange diamonds), percolating cubic diamond gel structures (pink to purple shaded circles), disordered percolating gel networks (blue triangles) and an isotropic fluid phase (light blue circles). Different shades of pink denote different number fractions of cubic diamond $N_D/N$ in the percolating structures following  the color bar.}
\label{figsi:no_dna}
\end{figure*}

%In the main text, we show the arise of diverse phases depending on the center-to-lobe size ratio $s$ and the lobe-to-depletant size ratio $q$ for systems where the particles in the core interact through a strong-short-range attractive interaction. 
To shed light on the effect of DNA on the self-assembly of cubic diamond, we show in Fig.~\ref{figsi:no_dna} the corresponding kinetic phase diagrams for tetrahedral colloidal molecules, but now without DNA functionalization of the central core particles. The topology of the kinetic phase diagrams resembles that of the ones in the main text for DNA-functionalized tetrahedral colloidal molecules. The kinetic phase diagrams exhibit a fluid phase at small depletant reservoir packing fractions $\eta_d^r$, corresponding to low interaction strengths $\beta U$ for all core-to-lobe size ratios $s$.  Upon increasing $\beta U$,  a narrow region is found in which cubic colloidal diamond is formed and  another small region where we observe the formation of percolating networks retaining an inner crystalline structure. These two phases are not observed for the largest core-to-lobe size ratio $s=1.28$. Finally, at larger interaction strengths $\beta U$ the system forms a percolating network. The latter is found in a larger region  compared to the model where the DNA interaction is considered. 
%As an example, at the largest center-to-lobe size ratios $s$ analyzed, that is $s=1.28$, the cubic diamond is not favorable anymore and only a fluid phase and a gel without crystalline structure is found. 
This, in general, reinforces the idea that the presence of DNA on the core particles  enhances and stabilizes the formation of cubic diamond crystals.

\section*{Probability distribution of the angular strain of six-membered rings}
\begin{figure}[h]
\includegraphics[width=0.5\linewidth]{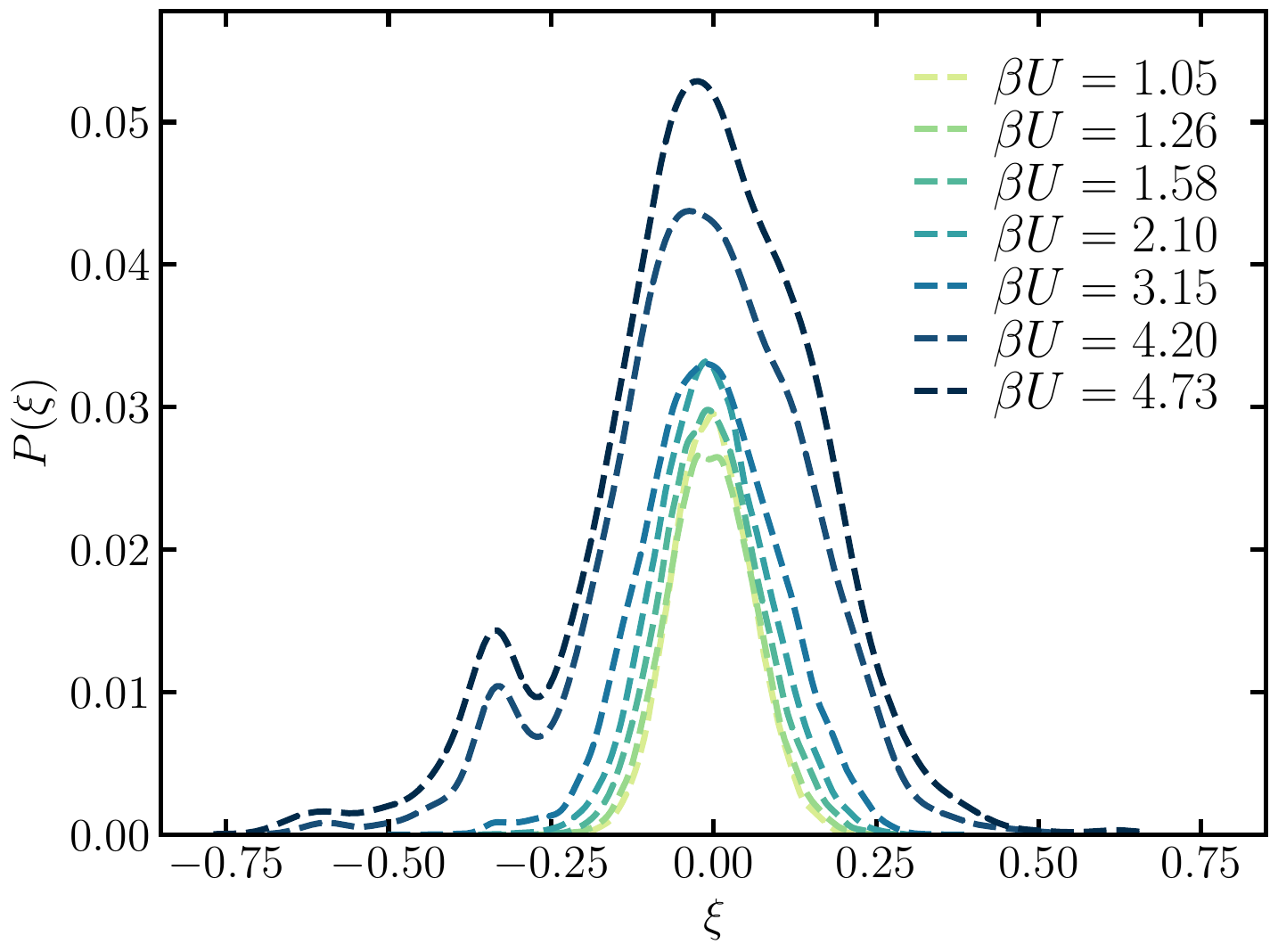} 
\caption{Probability distribution
of the angular strain $P(\xi)$ of six-membered rings for  different phases of DNA-functionalized tetrahedral colloidal molecules with a core-to-lobe size ratio $s=1.24$,  a depletant-to-lobe size ratio  $q=0.12$, and varying interaction strengths $\beta U$ as labeled.}
\label{figsi:strain}
\end{figure}

%As the depletion concentration $\eta_d^r$ increases, the strength of the attractive depletion interaction also increases, this, as it has been shown in the main text, leads to the formation of arrested phases. In particular, we find that at intermediate interaction strenghts there is an arise of percolating networks that retain a crystalline structure in their inside. As the attraction is increased the crystallinity of the percolating networks is lost and just random gel-like networks are found. We observe that as a consequence of the interlocking mechanism between the tetrahedral-particles in all the phases we find that the particles form with their nearest neighbors $6$-member rings. When the particles belong to a cubic diamond, all these $6$-member rings are in a chair-like conformation, where the innner angle between the particles is $\approx 109.5$. If the ring gets distorted deviating from the stable conformation, an angular strain is induced. We calculate the latter as $\theta_{ijk}-109.1$, where $jk$ are neighboring particles of particle $i$. 
In Fig.~\ref{figsi:strain}, we show the probability distribution $P(\xi)$ of the angular strain of six-membered rings for a system of DNA-functionalized tetrahedral colloidal molecules with a core-to-lobe size ratio $s=1.24$, a depletant-to-lobe size ratio $q=0.12$, and at different values of the interaction strengths $\beta U$ as labelled, corresponding to varying depletant reservoir packing fractions  $\eta_d^r$. At low $\beta U$, the system retains a cubic diamond structure and hence, most of the rings are in a chair-like or staggered conformation. As the interaction strength   $\beta U$ increases, the system gets arrested and the probability distribution $P(\xi)$ broadens accordingly. At high $\beta U$, the system fully looses crystallinity and a second peak in the distribution arises at a strain value around $-0.35$. 

\section*{Fractal dimension of the gel-like structures}
\begin{figure}
\includegraphics[width=0.5\linewidth]{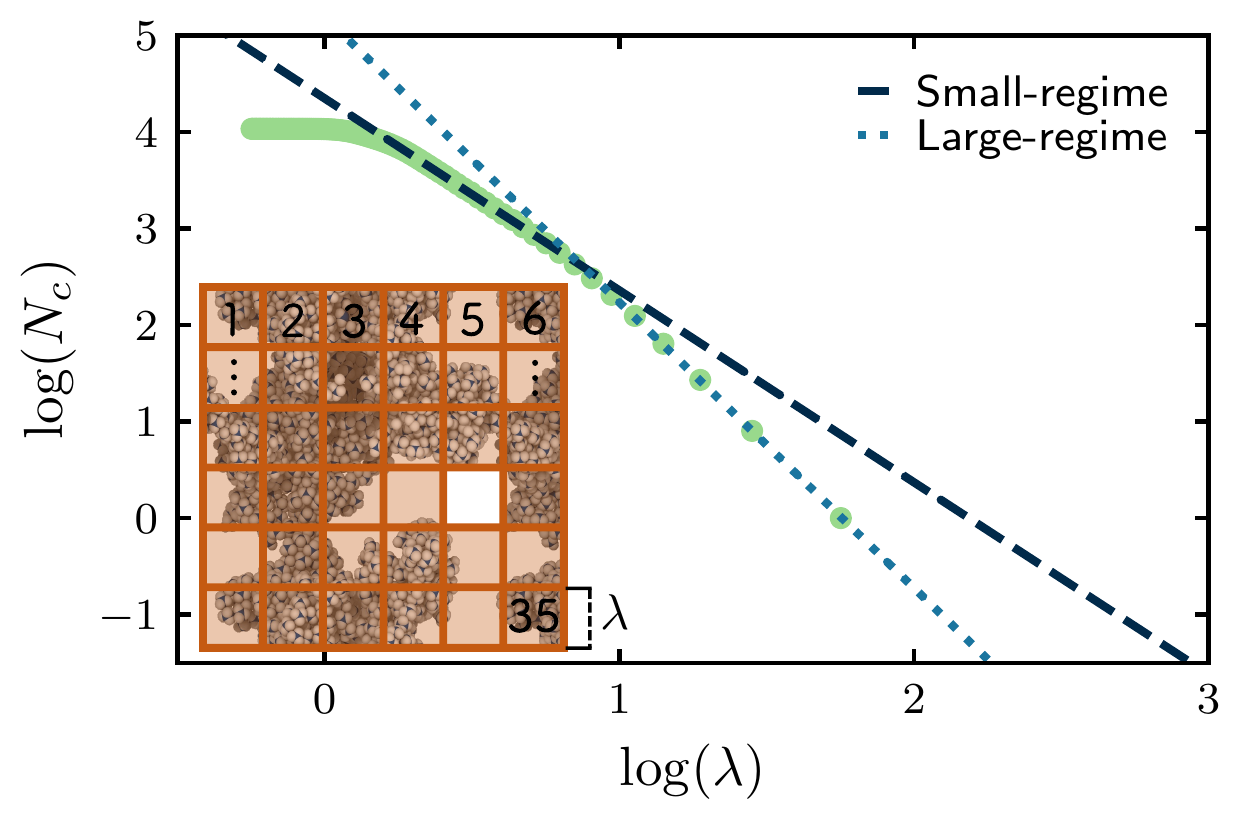} 
\caption{The logarithm of the number of cells $N_c$ with side length $\lambda$ that contain at least one colloidal particle as a function of $\lambda$. The dashed line and dotted line show the fits at small and large length scales usinge Eq. \ref{fit} for obtaining the fractal dimension $d_f$ for a gel structure of tetrahedral colloidal molecules with a core-to-lobe size ratio $s=1.24$, depletant-to-lobe size ratio $q=0.25$, depletant reservoir packing fraction $\eta_d^r=0.20$, and at colloid density  $\rho \sigma_A^3=0.06$. The inset shows a schematic of the box counting algorithm in which the simulation box is divided into small cells of side $\lambda$.}
%Two regimes of fractal dimension of a gel-like phase corresponding to the system of tetrahedral particles with $s=1.24$, $q=0.25$, $\eta_d^r=0.20$, $\rho=0.06$. Inset shows an example of the box counting algorithm.}
\label{figsi:fd_fit}
\end{figure}

In order to further characterize the nature of the percolating networks, we calculate the fractal dimension $d_f$ of our gel-like structures using the box counting algorithm. To this end, we first subdivide the simulation box into cells of varying size $\lambda$, and we count the number of cells $N_c$ that contain at least one colloidal particle. We show an example of this procedure in the inset of Fig.~\ref{figsi:fd_fit}. The fractal dimension $d_f$ is obtained using 
\begin{equation}
\log N_c(\lambda)=-d_f \log \lambda.   
\label{fit}
\end{equation}
we discriminate between two regimes by taking a cross-over value  $\lambda=8\sigma_A$, and use Eq. \ref{fit} to fit the two regimes. In Fig.~\ref{figsi:fd_fit} we show an example of such a fitting  for a percolating system of tetrahedral colloidal molecules with a core-to-lobe size ratio $s=1.24$, depletant-to-lobe size ratio $q=0.25$, and depletant reservoir packing fraction $\eta_d^r=0.20$.

\section*{Phase diagram for $q=0.25$}
\begin{figure}
\includegraphics[width=0.5\linewidth]{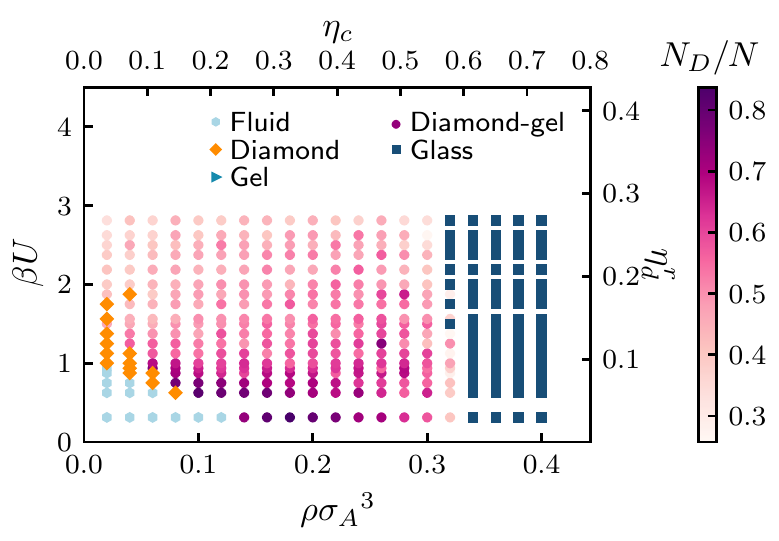} 
\caption{Kinetic phase diagram of  
a system of DNA-functionalized tetrahedral colloidal molecules with a core-to-lobe size ratio $s=\sigma_A/\sigma_B=1.24$ and a  depletant-to-lobe size ratio $q=\sigma_d/\sigma_A=0.25$ as a function of interaction strength $\beta U$ (left axis) or depletant reservoir packing fraction $\eta_d^r$ (right axis), and number density $\rho \sigma_A^3$ (lower axis),  or colloid packing fraction $\eta_c$ (upper axis). The color bar on the right indicates the fraction of particles in a diamond environment $N_D/N$.
}
\label{figsi:PD}
\end{figure}

 In Fig.~\ref{figsi:PD} we show the kinetic phase diagram for a system of DNA-functionalized tetrahedral colloidal molecules with a core-to-lobe size ratio $s=\sigma_B/\sigma_A=1.24$ and depletant-to-lobe size ratio   $q=\sigma_d/\sigma_A=0.25$ as a function of interaction strength $\beta U$ or depletant reservoir packing fraction $\eta_d^r$ and colloid number density $\rho\sigma_A^3$ or colloid packing fraction $\eta_c$. The phases that we found are consistent with the ones found for a system with smaller depletant size ratio $q=0.12$ shown in the main text.
 
\section*{Typical configurations as obtained from simulations  for high-density phases}
\begin{figure}
\includegraphics[width=0.85\linewidth]{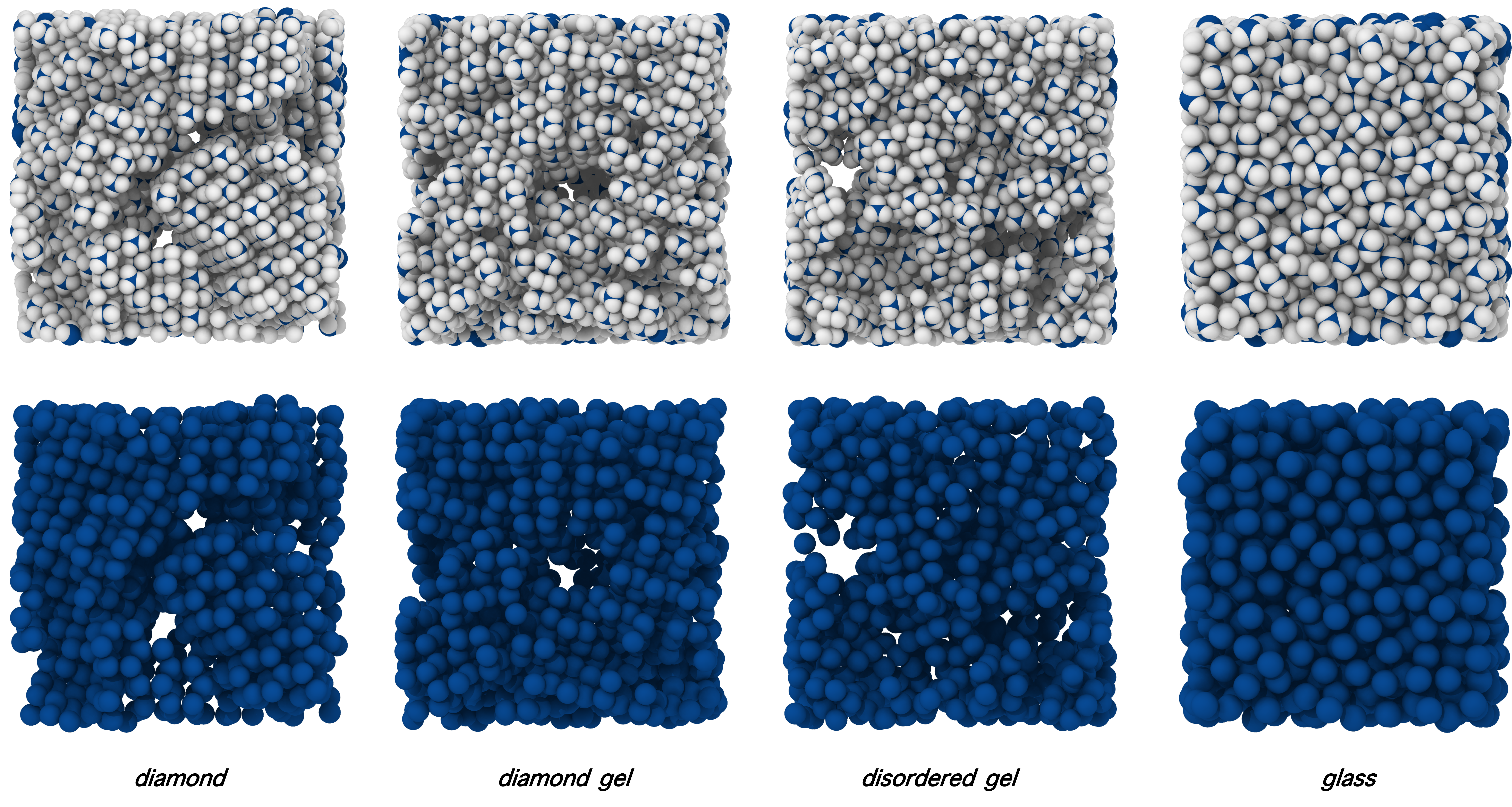}
\caption{Typical configurations obtained from simulations for  phases found for  tetrahedral colloidal molecules for a core-to-lobe size ratio $s=\sigma_B/\sigma_A=1.24$ and depletant-to-lobe size ratio $q=\sigma_d/\sigma_A=0.12$:  cubic diamond, diamond gel-like network,  disordered gel at a colloid number density $\rho \sigma_A^3=0.2$, and a glass at a colloid number density $\rho \sigma_A^3=0.4$ (from left to right). Top panels report the full colloidal molecules, while bottom panels display only the core particles.}
\label{figsi:snap}
\end{figure}

In Fig.~\ref{figsi:snap} we show representative configurations obtained from simulations for the phases observed for DNA-functionalized tetrahedral colloidal molecules at high  colloid number densities $\rho\sigma_A^3$.

\newpage
\section*{Characterization of the arrested phases}
\begin{figure*}[b!]
\includegraphics[width=0.7\linewidth]{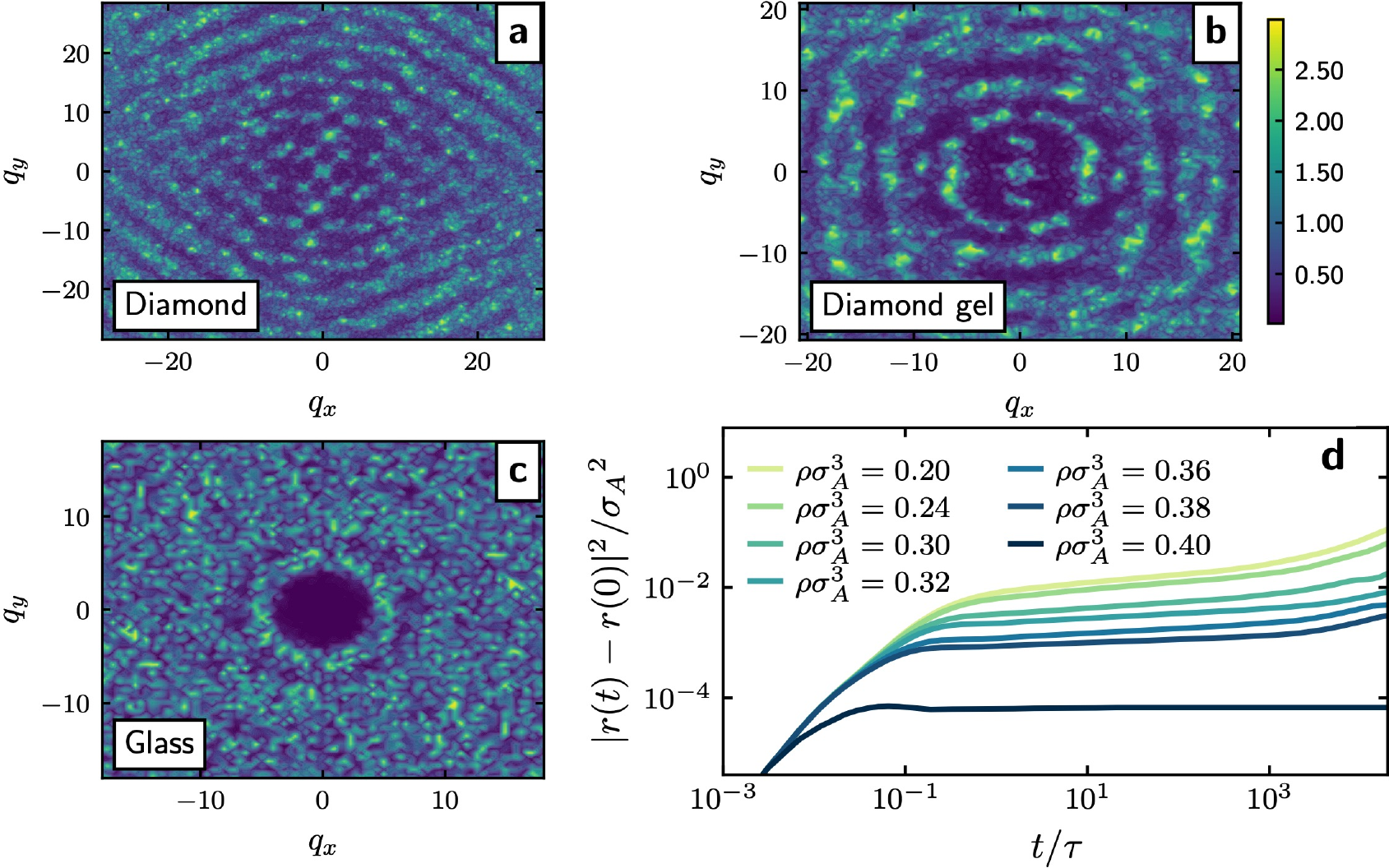} 
\caption{Diffraction patterns of the different phases for a system of DNA-functionalized tetrahedral colloidal molecules with a core-to-lobe size ratio $s=1.24$, a depletant-to-lobe size ratio $q=0.12$, and depletant reservoir packing fraction $\eta_d^r=0.12$ a) diamond phase at $\rho \sigma_A^3=0.06$, b) diamond gel-like structure at $\rho \sigma_A^3=0.20$ and c) a glass phase at $\rho \sigma_A^3=0.40$, and d) the mean-squared displacement of the respective system for varying colloid densities as labeled.}
\label{figsi:msd}
\end{figure*}
In the regime of high colloid packing fractions $\eta_c$ we find a region where the system is completely arrested and present no long-range order. In order to characterize these phases and their dynamical characteristics, we calculate the diffraction patterns and the mean squared displacement at high colloid number densities $\rho \sigma_A^3$. We show the results in Fig.~\ref{figsi:msd}(a-d). 

\section*{Crystal Nucleation of colloidal diamond}

\begin{figure}[h!]
\includegraphics[width=0.5\linewidth]{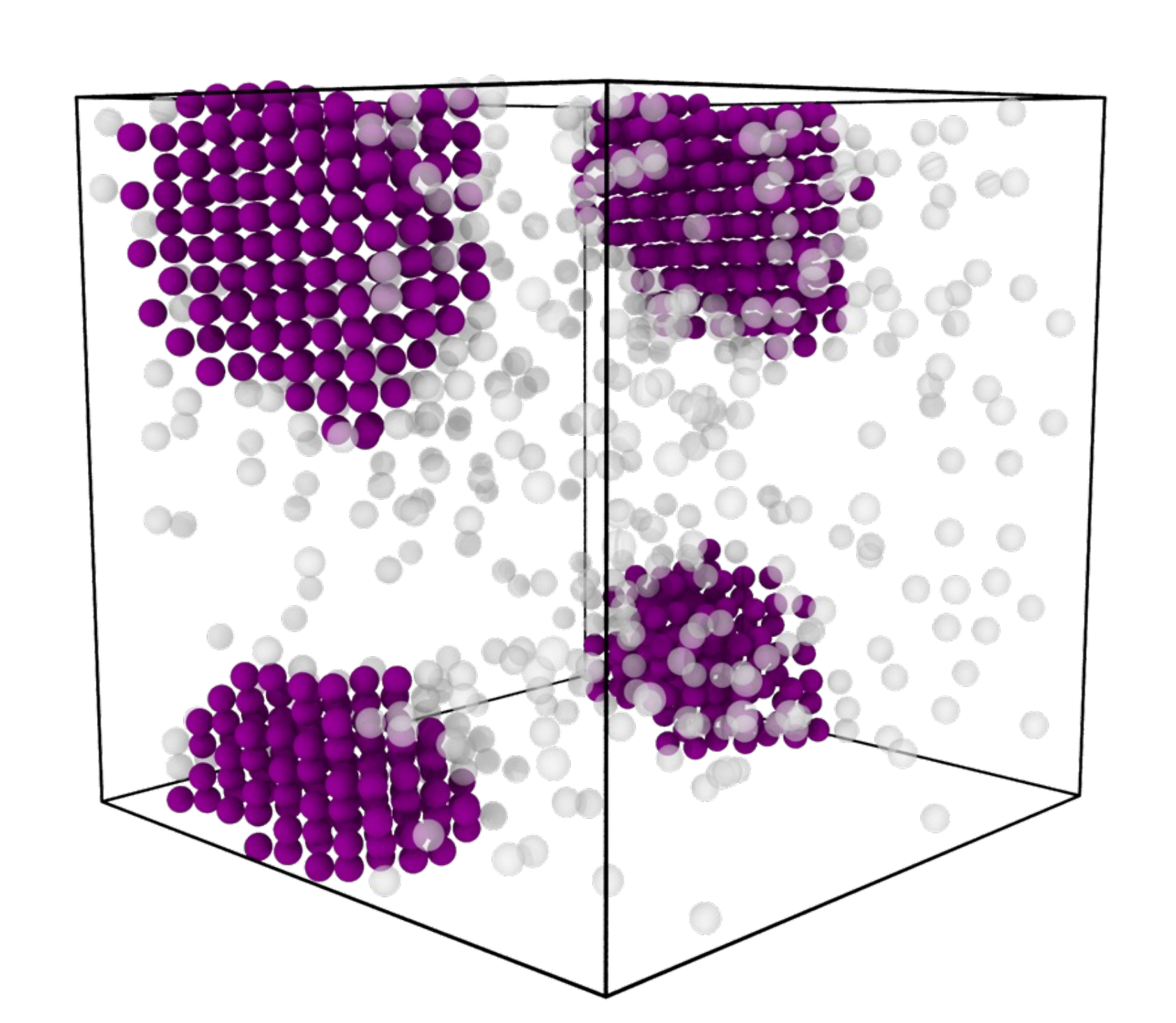} 
\caption{Frame of Video S1 showing the crystalline diamond grain of a system with a core-to-lobe size ratio $s=1.24$, depletant-to-lobe size ratio $q=0.12$, at a depletant reservoir packing fraction $\eta_d^r=0.10$. (Multimedia view)}
\label{figsi:nucleation}
\end{figure}

In the attached video (Video S1) we report for a representative case the nucleation of a crystalline diamond grain corresponding to a system with core-to-lobe size ratio $s=1.24$, depletant-to-lobe size ratio $q=0.12$, and at a depletant reservoir packing fraction $\eta_d^r=0.10$. Particles belonging to the growing cubic diamond grain are colored in purple. A frame of Video S1 is reported in Fig.~\ref{figsi:nucleation}.

\clearpage
\newpage

\twocolumngrid

\section*{References}
\bibliographystyle{apsrev4-1}
\bibliography{references}

\end{document}